\documentclass[12pt]{article}

\usepackage{amsmath,amssymb,graphics,graphicx,subfig,epsfig,wasysym}
\usepackage{xcolor,color,soul}

\newcommand\boxedB[1]{{\setlength\fboxsep{10pt}\boxed{#1}}}

\newcommand{\keywords}[1]{\textbf{Keywords:}\quad #1}

\begin{document}

\title{\bf Quantitative description of the surface tension minimum in a two-component surfactant system}

\author{Edgar M. Blokhuis}

\date{\it Colloid and Interface Science, Leiden Institute of Chemistry,\\ Leiden University, The Netherlands}

\maketitle

\begin{abstract}
\noindent
The Gibbs adsorption equation is the thermodynamic cornerstone for the description and understanding of the surface tension in a surfactant solution. It relates the decrease in surface tension to an increased surfactant adsorption. In the early 1940's, it therefore puzzled researchers to sometimes observe a {\em minimum} in the surface tension for certain surfactant solutions, which seemed to indicate surfactant desorption (even depletion) according to the Gibbs adsorption equation. It is now understood that the minimum is related to contamination of the surfactant (notably by dodecanol) and its occurrence has since then been studied extensively in experiments. Still, the precise role of the (tiny amount of) contaminant present is not well understood and a quantitative description and understanding of the minimum in the experimental surface tension is lacking. It is the aim of the present article to provide such a quantitative description. Our theoretical analysis is based on a Statistical Thermodynamic treatment of the Langmuir model for a surfactant mixture combined with the mass action model adapted to describe the formation of mixed micelles. A new Statistical Thermodynamic expression for the surface tension is derived and used to compare with a number of surface tension experiments for both ionic and non-ionic surfactant systems.
\end{abstract}

\noindent

\newpage
\section{Introduction}

It is well-known that the addition of a surfactant to a liquid leads to a decrease in the value of the liquid's surface tension. This phenomenon forms the basis of several oil recovery schemes and has (therefore) been the subject of extensive experimental, simulational and theoretical work. Key in understanding this phenomenon is the Gibbs adsorption equation \cite{Gibbs} which relates the slope of the decrease in surface tension to the amount of surfactant adsorbed at the interface. The decrease in surface tension with surfactant concentration is ultimately halted at the critical micelle concentration (cmc) when the surfactants in the bulk liquid start to reorganize themselves in micelles.

In the 1940's, it was noted that for some surfactant systems (labeled as type III systems \cite{McBain_1940a, McBain_1940b}), the surface tension would exhibit a clear {\bf minimum} in the vicinity of the cmc when plotted as a function of surfactant concentration \cite{McBain_1940a, McBain_1940b, Miles_1944}. This seemed in contradiction with the Gibbs adsorption equation which says that a rise in surface tension would indicate {\bf desorption} of the surfactant, which seemed quite unlikely. It was ultimately concluded that the minimum in the surface tension must be due to the presence of a tiny amount of a second, surface active component -- a contaminant. This conclusion was experimentally supported by showing that rigorous and repeated purification of the surfactant would ultimately lead to the expected monotonous decrease of the surface tension \cite{Miles_1944}. It was also concluded that the contamination is most likely due by the presence of traces of dodecanol (also known as lauryl alcohol or DOH) involved in the surfactant synthesis \cite{Miles_1944, Lin_1999}. Dodecanol molecules are highly surface active but they do not form micelles on their own. Since then, a number of experiments have been carried out that systematically investigate the occurrence and magnitude of the minimum in the surface tension by controlled addition of a second component, such as dodecanol, to a purified surfactant system \cite{Lin_1999, Vollhardt_2000, Denkov_2004, Razavi_2022}.

The physical picture that emerged to explain the minimum in the surface tension was subsequently formulated by several researchers \cite{Miles_1945, Crisp_discussion_1947, Reichenberg_1947, Brady_1949, Mysels_note_1996, Berg_book_2010}. At concentrations below the cmc, the presence of the contaminant lowers the surface tension because of the fact that its surface activity is higher than that of the surfactant. At a concentration somewhat lower than the cmc of the pure surfactant, {\em mixed} micelles (or pre-micelles) form that are composed of surfactant and contaminant. Due to the formation of these mixed micelles, the contaminant desorbs from the surface leading to the observed rise in surface tension.

Even though this explanation is not subject to contention, it proved difficult to describe the minimum in the surface tension in a {\em quantitative} manner. Furthermore, questions remain on the composition and size of the mixed micelles formed and their evolution as a function of surfactant concentration especially nearing the cmc of the pure surfactant solution. It is the aim of the present article to arrive at such a quantitative description of the full shape of the surface tension as a function of surfactant concentration in the presence of a certain amount of contaminant and address these questions on (the evolution of) the micellar composition. Our theoretical analysis is based on a Statistical Thermodynamic treatment of {\bf the Langmuir model} \cite{Langmuir, Rosen_book} extended to two surfactant types and {\bf the mass action model} \cite{Rusanov_book_1993} to describe (mixed) micellar formation.

Even though these two ingredients of our theoretical treatment are well-known \cite{Rusanov_book_1993, Gracia-Fadrique_2016} for the single surfactant system, we arrive at a new Statistical Thermodynamic expression for the surface tension of a mixture which proves to be especially useful to analyse the minimum in the surface tension. An important new element in this expression for the surface tension is that it takes into account the possible difference in (molar) surface areas of the two components. We compare our theoretical treatment to a number of surface tension experiments for both non-ionic (Section 2) and ionic (Section 3) surfactant solutions.

\section{Non-ionic surfactants}

\noindent
We consider a liquid-vapour system at fixed temperature $T$ to which surfactant molecules are added with a certain (number) concentration $c_{\rm s}$. We first consider a surfactant solution containing only a single, non-ionic type of surfactant.

\subsection{Single surfactant type}

\noindent
The surfactant molecule is considered to be present in solution in either one of two possible states: it is present as a monomer or it is part of a micelle. The total surfactant concentration is than the sum of the concentrations of surfactant monomers ($c_1$) and those part of a micelle ($c_{\rm m}$):
\begin{equation}
c_{\rm s} = c_1 + c_{\rm m} \,.
\end{equation}
In writing this equation, we have assumed that the bulk region is large enough to be able to neglect the amount of surfactant molecules adsorbed to the liquid surface (something which may not always be the case experimentally \cite{Fainerman_2017}).

We shall further assume that the chemical potential of the surfactant monomers corresponds to that of an {\em infinitely dilute} solution:
\begin{equation}
\label{eq:chemical_potential}
\mu_{\rm s} = \mu^{\circ}_{\rm s} + k_{\rm B} T \, \ln(c_1 / c^{\circ}) \,,
\end{equation}
where $\mu^{\circ}_{\rm s}$ is (by definition) the chemical potential at a reference concentration $c^{\circ}$ assuming that the solution is infinitely dilute. The reference chemical potential $\mu^{\circ}_{\rm s}$ depends on temperature, type of solvent and on the type of surfactant. 

Different choices for the reference concentration are possible and have been made in the literature. One common choice (that we will adopt here) is to relate $c^{\circ}$ to the (molar) volume $v_0$ of water \cite{Rosen_book}:
\begin{equation}
c^{\circ} = 1 / v_0 = {\rm 55.3 \,\, mol / L} \,.
\end{equation}
Central in our analysis is the evolution of the surface tension $\sigma$ of the liquid-vapor interface as surfactant is added to the solution. The change in surface tension is thermodynamically linked to the amount of surfactant adsorbed $\Gamma$ via the Gibbs adsorption equation \cite{Gibbs}:
\begin{equation}
\label{eq:Gibbs_adsorption_equation}
\left( \frac{\partial \sigma}{\partial \mu_{\rm s}} \right)_T = - \Gamma \,.
\end{equation}
To model surfactant adsorption, we shall use {\bf the Langmuir model} \cite{Rosen_book} in which the liquid-vapor interface is described in terms of a (fixed) number of adsorption sites that are available to the surfactant molecules (see also the Supporting Information). An important parameter is then the {\em adsorption energy} $\Delta E_{\rm s}$ associated with the adsorption of a surfactant molecule from a (reference) bulk solution.

In the Langmuir model, the adsorption is given by the well-known {\em Langmuir isotherm}
\begin{equation}
\label{eq:Langmuir_isotherm}
\frac{\Gamma}{\Gamma_{\infty}} = \frac{x}{1 + x} \,,
\end{equation}
where the parameter $x$ is defined as
\begin{equation}
\label{eq:x_definition}
x \equiv \exp \, [ \, (\mu_{\rm s} - \mu^{\circ}_{\rm s} - \Delta E_{\rm s}) / k_{\rm B} T \, ] \,.
\end{equation}
Furthermore, the Langmuir model leads to the following expression for the surface tension 
\begin{equation}
\label{eq:sigma_x}
\sigma = \sigma_0 - k_{\rm B} T \,\, \Gamma_{\infty} \, \ln(1 + x) \,,
\end{equation}
where $\sigma_0$ is the bare surface tension in the absence of surfactant.

It is convenient to relate $x$ to the surfactant monomer concentration $c_1$. Inserting the expression for $\mu_{\rm s}$ in Eq.(\ref{eq:chemical_potential}) into the definition for $x$ gives:
\begin{equation}
x = K \, c_1 \,,
\end{equation}
with
\begin{equation}
\label{eq:K}
K = v_0 \, \exp \, [ \, - \Delta E_{\rm s} / k_{\rm B} T \, ] \,.
\end{equation}
We then have that 
\begin{equation}
\label{eq:sigma}
\sigma = \sigma_0 - k_{\rm B} T \,\, \Gamma_{\infty} \, \ln(1 + K \, c_1) \,.
\end{equation}
This expression for the surface tension describes the behaviour of the surface tension in the entire concentration regime below {\em and} above the cmc. However, it is then necessary to relate the concentration of monomers $c_1$ to the overall surfactant concentration $c_{\rm s} \!=\! c_1 + c_{\rm m}$. In other words, we need to theoretically model micelle formation. Before doing so, we first investigate this expression for $\sigma$ in the {\em dilute regime} prior to micelle formation ($c_{\rm m} \!\approx\! 0$), so that we can approximate $c_1 \!\approx\! c_{\rm s}$.

\subsubsection{Single surfactant type -- dilute regime}

\begin{figure}[ht]
\centering
\subfloat[dilute regime]{\includegraphics[width=0.49\textwidth]{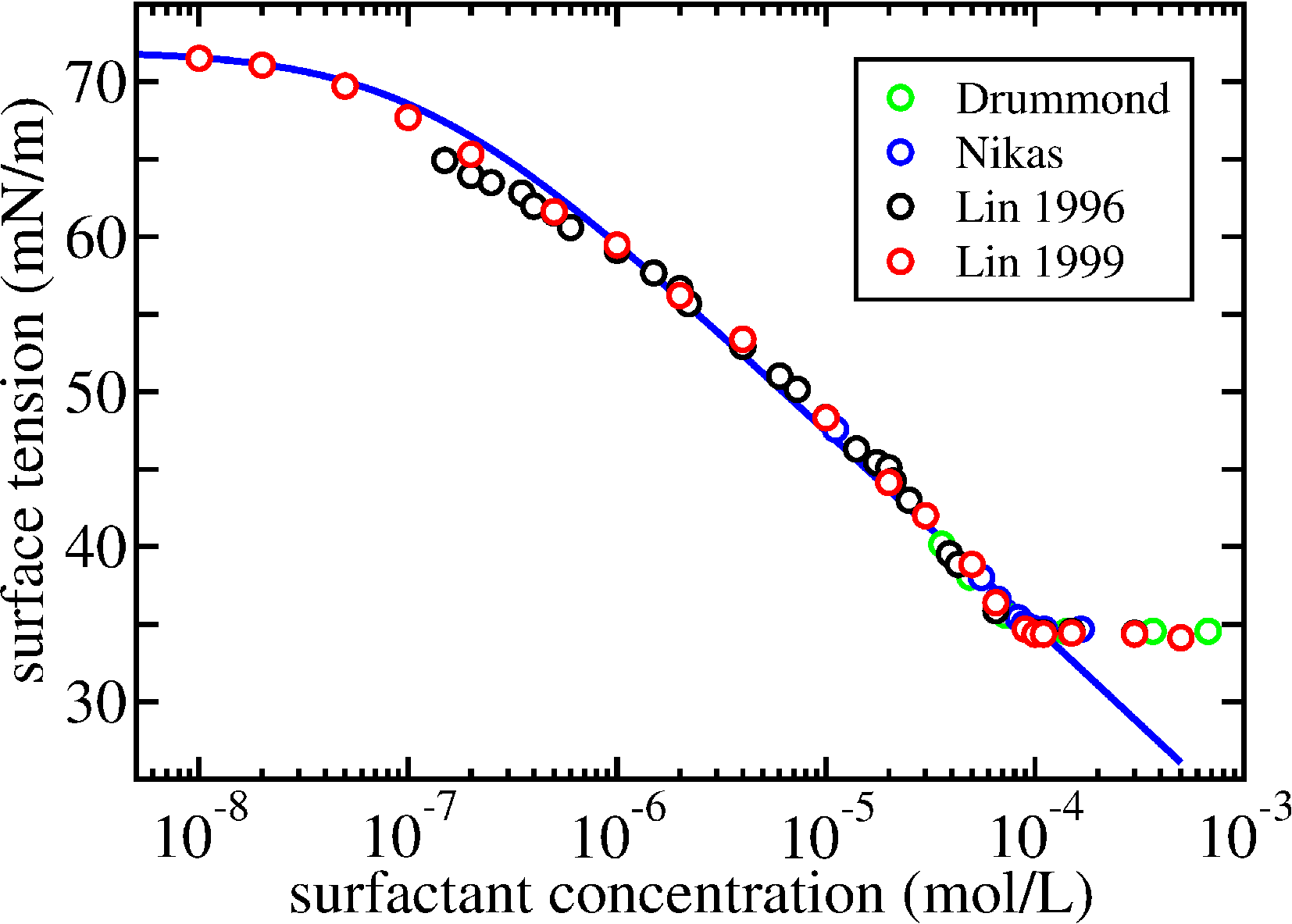}
\label{Fig:C12E8_single}}
\hfill
\subfloat[micelle formation]{\includegraphics[width=0.49\textwidth]{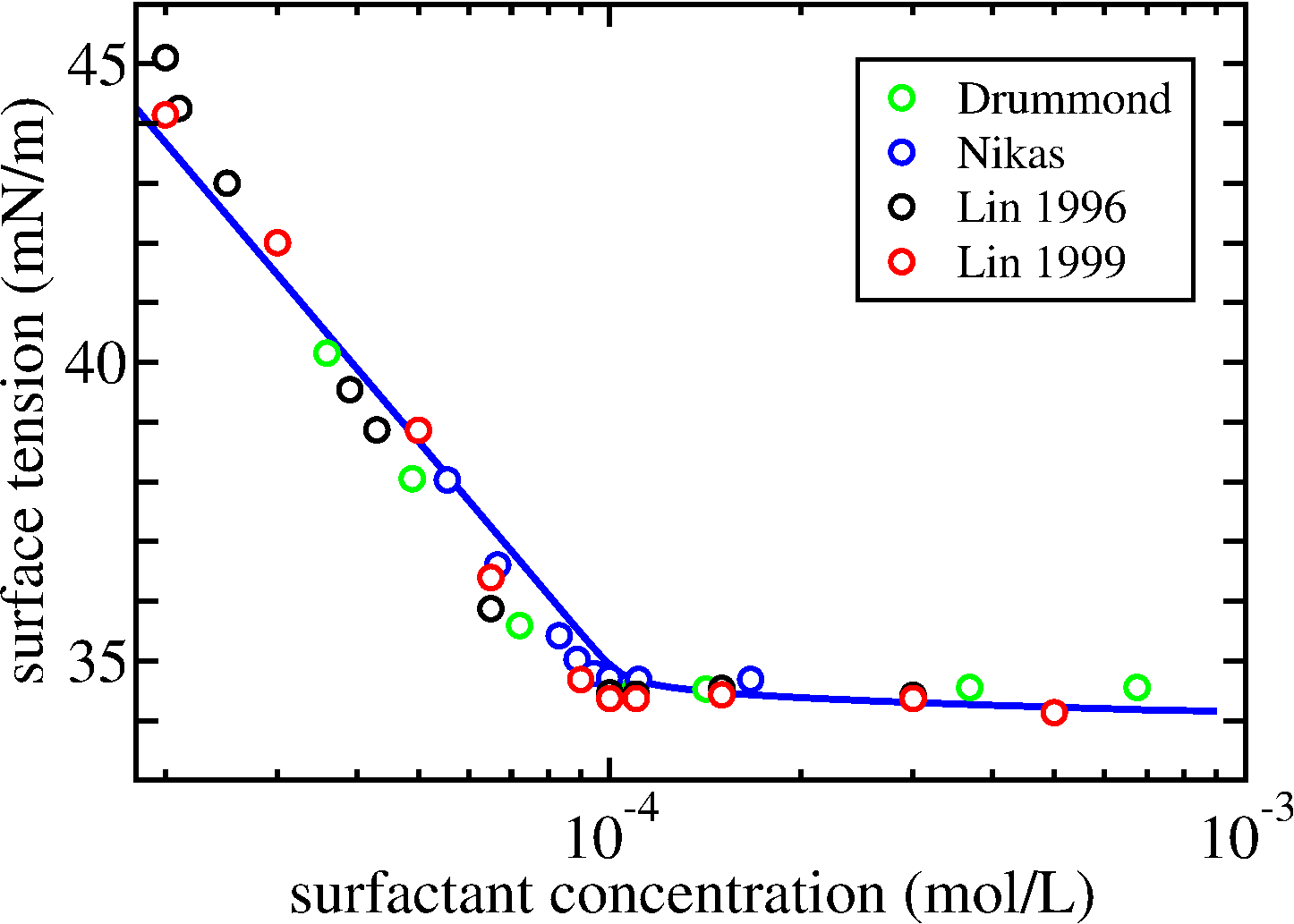}
\label{Fig:C12E8_single_micel}}
\caption{Surface tension as a function of surfactant concentration for an aqueous solution of C$_{12}$E$_8$. Open symbols are experimental results by Drummond {\em et al.} \cite{Ninham_1985}, Nikas {\em et al.} \cite{Nikas_1992} and Lin {\em et al.} \cite{Lin_1999, Lin_1996}. The results in refs.\cite{Ninham_1985} and \cite{Nikas_1992} are shifted by 1.7 mN/m to account for a somewhat different bare surface tension which is taken to be $\sigma_0 \!=$ 72.0 mN/m. In (a), the solid line is the Langmuir-Szyskowski equation in Eq.(\ref{eq:sigma_LS}) with values for the fit parameters $\Gamma_{\infty}$ and $K$ listed in Table \ref{Table:data_single}. In (b), the solid line is Eq.(\ref{eq:sigma}) with values listed in Table \ref{Table:data_single} for the additional fit parameters $m$ and $x_0$ from the mass action model in Section \ref{single_surfactant_micelle_formation}.}
\label{Fig:Figure_1}
\end{figure}

\noindent
In the dilute regime, the expression for the surface tension in Eq.(\ref{eq:sigma}) reduces to the so-called {\em Langmuir-Szyszkowski} equation \cite{Rosen_book}
\begin{equation}
\label{eq:sigma_LS}
\sigma \approx \sigma_0 - k_{\rm B} T \,\, \Gamma_{\infty} \, \ln(1 + K \, c_{\rm s})  \hspace*{50pt} {\rm (dilute)}
\end{equation}
The surface tension as a function of surfactant concentration in this regime is thus described in terms of two fit parameters $\Gamma_{\infty}$ and $K$. For most surfactant systems the experimentally measured surface tension for dilute solutions is rather well described by the Langmuir-Szyszkowski expression in Eq.(\ref{eq:sigma_LS}) up to the cmc thus providing experimental estimates for $\Gamma_{\infty}$ and $K$ \cite{Abott_book}. A typical example is shown in Figure~\ref{Fig:C12E8_single} where the surface tension of an aqueous solution of C$_{12}$E$_8$ against air is shown at room temperature ($T \!=$ 298 K).

%
%

\noindent
Figure~\ref{Fig:C12E8_single} shows that up to a certain concentration, the surface tension is well described by the two fit parameters: the inverse of $K$ is the concentration at which the curve crosses over from the constant surface tension $\sigma_0$ of the water-air interface to the constant slope $\Gamma_{\infty}$ at intermediate concentrations. Since the Langmuir model only takes the repulsive interactions between surfactant molecules at the surface into account, it does not account for possible attractive interactions at larger separations. The inclusion of such attractive interactions in the form of a van der Waals-like surface equation of state (EOS) or via the phenomenological Frumkin model \cite{Rosen_book, Frumkin} would improve the agreement with experimental data for the adsorption and surface tension (see e.g. refs.\cite{Kralchevsky_1999, Kralchevsky_2002, Kralchevsky_2003}), but would do so at the expense of the introduction of another (interaction) fit parameter \cite{Nguyen_2018}.

Figure~\ref{Fig:C12E8_single} also shows that the Langmuir-Szyszkowski equation breaks down beyond the cmc due to the formation of micelles.

%
%

\begin{table}
\centering
\begin{tabular}{c || c | c | c || c | c | c }
& $\Gamma_{\infty}$ & $1/K$ & $\Delta E_{\rm s}$ & \hspace*{1pt} $m$ \hspace*{1pt} & $x_0$ & $\Delta E_{\rm m}$ \\
& \hspace*{1pt} 10$^{-6}$ mol/m$^2$ \hspace*{1pt} & \hspace*{1pt} 10$^{-7}$ mol/L \hspace*{1pt} & \hspace*{1pt} kJ/mol \hspace*{1pt} & & \hspace*{1pt} 10$^{-6}$ \hspace*{1pt} & \hspace*{1pt} kJ/mol \hspace*{1pt} \\
\hline
\hline
C$_{12}$E$_8$ & 2.21 \cite{Lin_1999} & 1.14 \cite{Lin_1999} & - 49.6 &  50 & 2.8  & - 31.7 \\
C$_{10}$E$_8$ & 2.04 \cite{Lin_1999} & 9.41 \cite{Lin_1999} & - 44.3 &  50 & 29.5 & - 25.8 \\
LSA           & 6.5                  & 11000                & - 26.8 & 100 & 165  & - 21.6 \\
\end{tabular}
\caption{Values of the fit parameters used to plot the theoretical curves in Figures~\ref{Fig:Figure_1}-\ref{Fig:Figure_3} of the purified surfactant system. The first two columns are the fit parameters $\Gamma_{\infty}$ and $K$ in the dilute concentration regime. The third column is the adsorption energy calculated from $K$ using Eq.(\ref{eq:K}) \cite{reference_state}. The fourth and fifth column are the fit parameters $m$ and $x_0$ to describe micelle formation. The final column is the energy gain for a surfactant to be part of a micelle calculated from $x_0$ using Eq.(\ref{eq:x_0}) \cite{reference_state}. The values of the fit parameters $\Gamma_{\infty}$ and $K$ for C$_{12}$E$_8$ and C$_{10}$E$_8$ are taken from Table 2 in ref.\cite{Lin_1999}.}
\label{Table:data_single}
\end{table}

\subsubsection{Single surfactant type -- micelle formation}
\label{single_surfactant_micelle_formation}

\noindent
Micelle formation will be described in terms of {\bf the mass action model} which has been widely used in this context \cite{Rusanov_book_1993}. In the minimum version of this model, the micelles are modeled in terms of two parameters: the (fixed) number $m$ of surfactant molecules that constitute the micelle and the energy gain of a single surfactant molecule when it becomes part of the micelle, $\Delta E_{\rm m}$. In the following, we discuss the consequences of the mass action model by considering the system's free energy. The approach via the free energy is especially useful if one wants to consider future adaptations of the mass action model to describe {\bf mixed} micelles.

The free energy in the mass action model is a function of the concentrations of surfactant monomers $c_1$ and those part of the micelles $c_{\rm m}$. It is, however, convenient to introduce (dimensionless) volume fractions $x_1 \!=\! v_0 \, c_1$ and $x_m \!=\! v_0 \, c_{\rm m}$. The free energy is then given in terms of $x_1$ and $x_m$ as
\begin{equation}
\frac{v_0}{V} \, F(x_1,x_m) = x_1 \, k_{\rm B} T \, (\ln(x_1) - 1 ) + \frac{x_m}{m} \, k_{\rm B} T \, (\ln(\frac{x_m}{m}) - 1) + x_m \, \Delta E_{\rm m} \,.
\end{equation}
The first two terms represent the translational entropy of surfactant monomers and of the micelles, respectively. The third term represents the energy gain of all surfactant molecules in the micelles. The free energy is to be minimized with respect to $x_1$ and $x_m$ under the constraint that $v_0 \, c_{\rm s} \!\equiv\! X_{\rm s} \!=\! x_1 + x_m$. It leads to the following expression for the surfactant volume fraction $x_m$
\begin{equation}
x_m = m \left( \frac{x_1}{x_0} \right)^{\!m} \,,
\end{equation}
where we have defined
\begin{equation}
\label{eq:x_0}
x_0 \equiv \exp \, [ \, \Delta E_{\rm m} / k_{\rm B} T \, ] \,.
\end{equation}
If one investigates the evolution of the volume fractions $x_1$ and $x_m$ as a function of the total surfactant concentration $X_{\rm s}$, a transition from mostly monomers to mostly micelles appears around $X_{\rm s} \!\approx\! x_0$. We can therefore interpret $x_0$ as the critical micelle concentration:
\begin{equation}
c_{\rm cmc} \approx x_0 / v_0 = c^{\circ} \exp \, [ \, \Delta E_{\rm m} / k_{\rm B} T \, ] \,.
\end{equation}
Since the mass action model predicts the surfactant monomer concentration to be more or less constant beyond the cmc \cite{Rusanov_book_1993, Rusanov_2014, Rusanov_2017}, it thus explains the leveling off of the surface tension at higher concentrations.

Using $m$ and $x_0$ as fit parameters, we show in Figure~\ref{Fig:C12E8_single_micel} the surface tension of an aqueous solution of C$_{12}$E$_8$ against air in the entire concentration regime. As discussed, the fit parameter $x_0$ is roughly equal to the concentration at which the surface tension levels off. It is a bit more difficult to directly relate the value of the micelle size $m$ to the characteristics of the experimental data. In general, the higher the value of $m$, the sharper is the transition at the cmc. Also, for lower values of $m$ the surface tension tends to slope downward a bit more beyond the cmc. In fact, it has been argued that the slope beyond the cmc could provide an accurate way to determine the (average) micellar size \cite{Gracia-Fadrique_2016}, although it has also been argued that care has to be taken with such an identification \cite{Rusanov_2017}.

Even though some crude approximations have been made, the overall agreement with experiment as shown in Figure~\ref{Fig:C12E8_single_micel} is rather satisfactory. Despite the fact that as many as four fit parameters are to be determined, they can all be obtained rather unambiguously from different concentration regions. Furthermore, it is well documented that even better agreement with experiment can be obtained by including a van der Waals type of interaction between the surfactants on the surface \cite{Rosen_book}. The purpose of this article is to come to a similar agreement using a bare minimum of additional fit parameters for a {\bf surfactant-contaminant mixture}. We are especially interested in the situation where the concentration of the second component is small ($< 1\%$) but where the effect on the surface tension is substantial due to its high surface activity.

\newpage
\subsection{Surfactant-contaminant mixture}

\noindent
We now extend the previous analysis to a mixture of a surfactant (species $a$) and a contaminant (species $b$) with concentrations $c_a$ and $c_b$. We shall assume that the contaminant is present in a certain mole fraction $\alpha$ of the first component, i.e. $c_b \!=\! \alpha \, c_a$. For the systems considered here $\alpha$ is small and may or may not be known experimentally. Furthermore, it turns out to be important to allow for the possibility that the surfactant and contaminant take up different surface areas on the surface due to a difference in size of the respective polar head groups. This difference in size is described by a parameter $\beta \equiv \Gamma_{b,\infty} / \Gamma_{a,\infty}$ which is larger than 1 when the dominant surfactant species ($a$) takes up more surface area than the contaminant ($b$). In the Appendix (see also the Supporting Information), it is shown that a Statistical Thermodynamic treatment of the Langmuir model extended to such a mixture leads to the following expression for the surface tension
\begin{equation}
\label{eq:sigma_mixture_x}
\sigma = \sigma_0 - k_{\rm B} T \,\, \Gamma_{a,\infty} \, \ln(x_a + (1 + x_b)^{\beta}) \,,
\end{equation}
where, analogously to before, $x_i$ ($i = a, b$) is defined as 
\begin{equation}
x_i \equiv \exp \, [ \, (\mu_i - \mu^{\circ}_i - \Delta E_{\rm s, i}) / k_{\rm B} T \, ] \,.
\end{equation}
It is shown in the Appendix that this expression for the surface tension is strictly derived under the condition that the parameter $\beta \geqslant$ 1, i.e. the contaminant is smaller than the dominant surfactant at the surface.

Again, as before, it is convenient to relate $x_i$ to the surfactant monomer concentration $c_{1,i}$. Inserting the expression for the chemical potential in Eq.(\ref{eq:chemical_potential}) into the definition for $x_i$ gives:
\begin{equation}
x_i = K_i \, c_{1, i}
\end{equation}
with
\begin{equation}
K_i = v_0 \, \exp \, [ \, - \Delta E_{\rm s, i} / k_{\rm B} T \, ] \,.
\end{equation}
We then have that 
\begin{equation}
\label{eq:sigma_mixture}
\boxedB{ \sigma = \sigma_0 - k_{\rm B} T \,\, \Gamma_{a,\infty} \, \ln(K_a \, c_{1, a} + (1 + K_b \, c_{1, b})^{\beta}) }
\end{equation}
This is the expression for the surface tension of a mixture of surfactant and contaminant that we shall use to compare to a number of experimental examples. One may verify that when the concentration of one of the components is zero, the expression reduces to the expression in Eq.(\ref{eq:sigma}) for a single component surfactant system, as it should. Furthermore, to be consistent, we shall always assume that surfactant specific quantities such as $\Gamma_{a, \infty}$ and $K_a$ are given by their values determined in the experiments for the single surfactant system, i.e. $K_a \!=\! K$ and $\Gamma_{a, \infty} \!=\! \Gamma_{\infty}$.

\subsubsection{Surfactant-contaminant mixture -- dilute regime}

\noindent
Again, we consider first the dilute regime so that we can approximate $c_{1, a} \!\approx\! c_{\rm s}$ and $c_{1, b} \!\approx\! \alpha \, c_{\rm s}$. The resulting approximate expression for the surface tension is the extension of the Langmuir-Szyszkowski equation to a two component mixture
\begin{equation}
\label{eq:sigma_mixture_LS}
\sigma \approx \sigma_0 - k_{\rm B} T \,\, \Gamma_{a,\infty} \, \ln(K_a \, c_{\rm s} + (1 + K_b \, \alpha \, c_{\rm s})^{\beta}) \hspace*{50pt} {\rm (dilute)}
\end{equation}

%
%

\begin{figure}[ht]
\centering
\subfloat[dilute regime]{\includegraphics[width=0.48\textwidth]{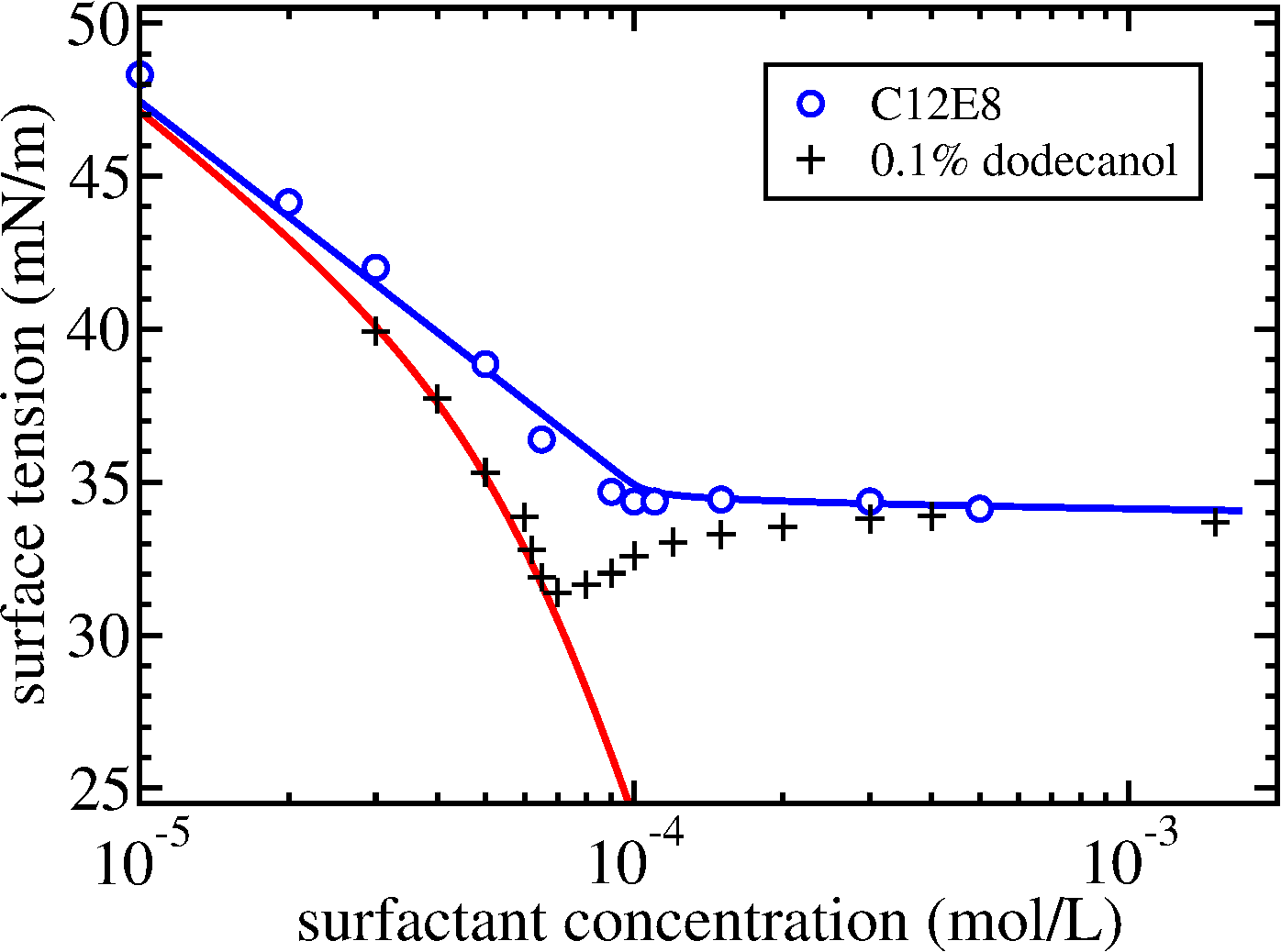}
\label{Fig:C12E8_mixture}}
\hfill
\subfloat[micelle formation]{\includegraphics[width=0.48\textwidth]{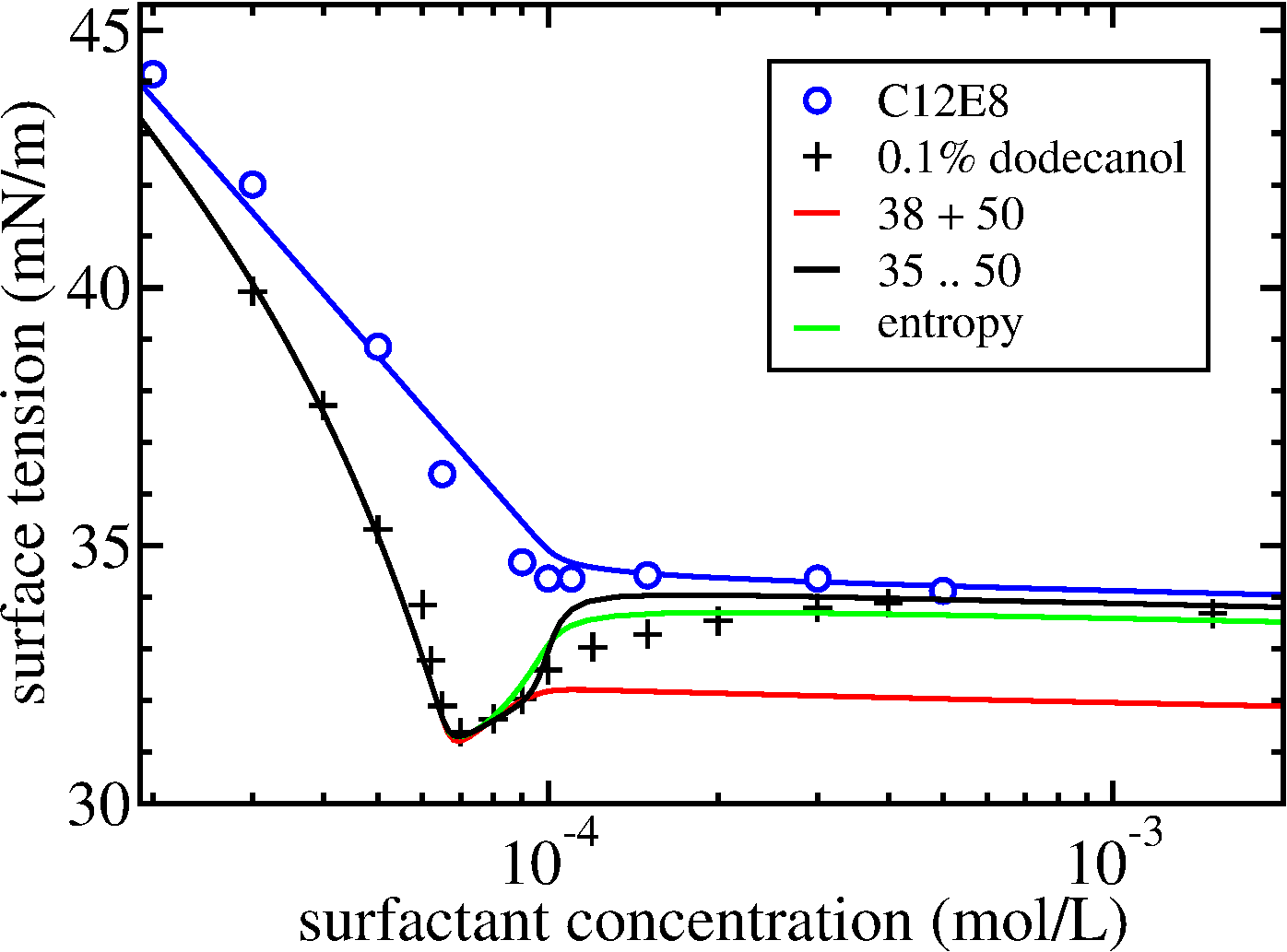}
\label{Fig:C12E8_mixture_micel}}
\caption{Surface tension as a function of surfactant concentration for an aqueous solution of C$_{12}$E$_8$. Symbols are the experimental results by Lin {\em et al.} \cite{Lin_1999} for the purified C$_{12}$E$_8$ system (open symbols) and with 0.10\% dodecanol added (plus signs). The solid blue line is the theoretical result in Figure~\ref{Fig:C12E8_single_micel} for the purified C$_{12}$E$_8$ system. In (a) the solid red line is Eq.(\ref{eq:sigma_mixture_LS}). In (b) the solid lines are the theoretical results in Section \ref{surfactant-contaminant_mixture_micelle_formation} for three different assumptions on the micellar composition: (1) solid red line: regular micelles ($m_a \!=$ 50) + mixed micelles ($m_a \!=$ 38); (2) solid black line: range of (mixed) micelle compositions ($m_a \!=$ 35 to $m_a \!=$ 50); (3) solid green line: no restriction on the composition range but with mixing entropy included. Values of the additional fit parameters are listed in Table \ref{Table:data_mixture}.}
\label{Fig:Figure_2}
\end{figure}

\noindent
In Figure~\ref{Fig:C12E8_mixture} this approximate expression for the surface tension is shown as the solid red curve together with experimental results by Lin {\em et al.} \cite{Lin_1999} for purified C$_{12}$E$_8$ and for purified C$_{12}$E$_8$ to which 0.10\% of dodecanol is added. With $\alpha \!=$ 0.10\% and with $\Gamma_{a,\infty} \!=\! \Gamma_{\infty}$ and $K_a \!=\! K$ determined earlier, the fit is obtained using only two fit parameters $\beta$ and $K_b$ (see Table~\ref{Table:data_mixture}). Furthermore, the fit value $\beta \!=$ 5 is consistent with independent measurements of the adsorption of the fully saturated pure C$_{12}$E$_8$ system (Table 2 in ref.\cite{Lin_1999}), $\Gamma_{a, \infty} \!=$ 2.21 $\times$ 10$^{-6}$ mol/m$^2$, and $\Gamma_{b, \infty} \!=$ 11.1 $\times$ 10$^{-6}$ mol/m$^2$ of pure dodecanol on water, leading to $\beta = \Gamma_{b, \infty}/\Gamma_{a, \infty} \!=$ 5.02.

Figure~\ref{Fig:C12E8_mixture} shows the considerable lowering of the surface tension due the presence of only a small amount of surface-active contaminant. Furthermore, the increasing downward slope indicates that a contaminant molecule has a smaller surface area than the surfactant molecule, i.e. $\beta \!>\! 1$. It is concluded that the experiments are well-described by the extended Langmuir-Szyszkowski equation in Eq.(\ref{eq:sigma_mixture_LS}) for low concentrations with, in this case, essentially only one additional fit parameter $K_b$. Again, it is also observed that the (extended) Langmuir-Szyszkowski equation breaks down at a certain concentration due to the formation of micelles.

%
%

\begin{table}
\centering
\begin{tabular}{c || c | c | c || c | c }
& \hspace*{1pt} $\alpha$ ($\%$) \hspace*{1pt} & \hspace*{5pt} $\beta$ \hspace*{5pt} & \hspace*{1pt} $K_b / K_a$ \hspace*{1pt} & composition & \hspace*{1pt} $x_0^{\rm b}$ (10$^{-9}$) \\
\hline
\hline
                          &                      &      &       & 38 + 50   & 1.15 \\
C$_{12}$E$_8$ + dodecanol & 0.10 \cite{Lin_1999} & 5    & 5.25  & 35 .. 50  & 1.20 \\
                          &                      &      &       & entropy   & 1.70 \\
\hline
                          &                      &      &       & 41 + 50   & 11.1 \\
C$_{10}$E$_8$ + dodecanol & 0.10 \cite{Lin_1999} & 1.75 & 120   & 38 .. 50  & 13.0 \\
                          &                      &      &       & entropy   & 39.0 \\
\hline
                          &                      &      &       & 30 + 100  & 14.8 \\
LSA + contaminant         & 0.04                 & 1    & 11375 & 30 .. 100 & 14.8 \\
                          &                      &      &       & entropy   & 32.5 \\
\end{tabular}
\caption{Values of the fit parameters used to plot the theoretical curves in Figures~\ref{Fig:Figure_2}-\ref{Fig:Figure_3} of the surfactant + contaminant system. The first two columns are the fit parameters $\alpha$, $\beta$ and $K_b$ in the dilute concentration regime. The amount of dodecanol ($\alpha \!=$ 0.10\%) for C$_{12}$E$_8$ and C$_{10}$E$_8$ is set by the experimental conditions in ref.\cite{Lin_1999}. The fourth column indicates which model is used to describe the micellar composition. The final column is the fit parameter $x_0^{\rm b}$ used to describe micelle formation.}
\label{Table:data_mixture}
\end{table}

\subsubsection{Surfactant-contaminant mixture -- micelle formation}
\label{surfactant-contaminant_mixture_micelle_formation}

In order to extend the mass action model to describe micelle formation in a two component surfactant mixture, we first need to consider the {\bf composition} of the micelles. We shall denote the number of molecules of the dominant species $a$ (the surfactant) in the micelle as $m_a$ and the number of molecules of species $b$ (the contaminant) as $m_b$. We shall allow $m_a$ to vary (see below) but always in such a way that the overall micellar size is the same as the uncontaminated system. This means that if the number of molecules of species $a$ in a micelle is less than $m$, all the freed up surface area of the micelle is filled by the smaller species $b$. This implies that
\begin{equation}
m_b = \beta \, (m - m_a) \,.
\end{equation}
Necessarily, when $m_a \!=\! m$, we have that $m_b \!=$ 0, and the micellar composition is that of the uncontaminated system. Furthermore, we shall denote the energy gain for each species $a$ and $b$ to be part of a micelle as $\Delta E_{\rm m, a}$ and $\Delta E_{\rm m, b}$, respectively.

As the system may comprise micelles of different compositions, the micellar free energy contribution is, in principal, a sum over all integer values between 0 and $m$ of the composition variable $m_a$. The free energy thus becomes a function of the two surfactant monomer volume fractions $x_1^a$ and $x_1^b$ and the {\em distributions} $\{x_m^a\}$ and $\{x_m^b\}$
\begin{eqnarray}
&& \frac{v_0}{V} \, F(x_1^a,\{x_m^a\},x_1^b,\{x_m^b\}) = x_1^a \, k_{\rm B} T \, (\ln(x^a_1) - 1) + x_1^b \, k_{\rm B} T \, (\ln(x^b_1) - 1) \nonumber \\
&& + \sum_{m_a} \left[ \frac{x_m^a}{m_a} \, k_{\rm B} T \, (\ln(\frac{x_m^a}{m_a}) - 1) + x_m^a \, \Delta E_{\rm m, a} + x_m^b \, \Delta E_{\rm m, b} \right] \,.
\end{eqnarray}
The first two terms denote the translational entropy of the surfactant monomers of both types. The last term is a summation over all values of $m_a$ of the free energy of micelles consisting of the micellar translational entropy and the energy gain. The free energy is to be minimized with respect to $x_1^a$, $x_1^b$ and the distributions $\{x_m^a\}$ and $\{x_m^b\}$ under the constraint that $x_1^a + \sum x_m^a \!=\! X_s$, $x_1^b + \sum x_m^b \!=\! \alpha X_s$ and $x_m^a / m_a \!=\! x_m^b / m_b$. The minimization leads to the following expression for the micellar composition distribution
\begin{equation}
\label{eq:composition_distribution}
\frac{x_m^a}{m_a} = \left( \frac{x_1^a}{x_0^a} \right)^{\!\!m_a} \left( \frac{x_1^b}{x_0^b} \right)^{\!\!m_b} \,,
\end{equation}
where $m_a$ runs over all values allowed and where we have defined ($i = a, b$)
\begin{equation}
x_0^i \equiv \exp \, [ \, \Delta E_{\rm m, i} / k_{\rm B} T \, ]  \,.
\end{equation}

\begin{figure}[ht]
\centering
\subfloat[C$_{10}$E$_8$]{\includegraphics[width=0.49\textwidth]{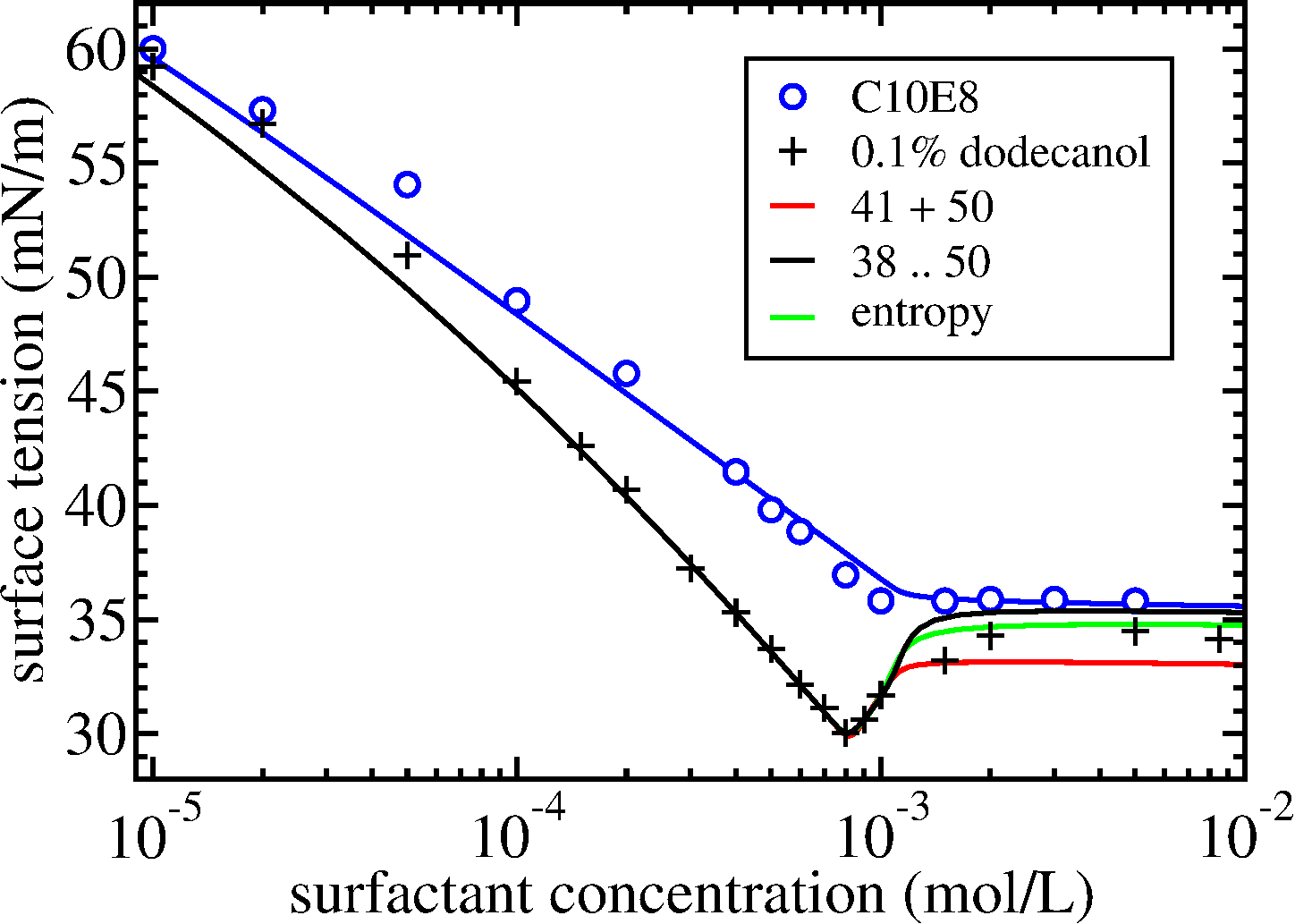}
\label{Fig:C10E8_mixture_micel}}
\hfill
\subfloat[LSA]{\includegraphics[width=0.49\textwidth]{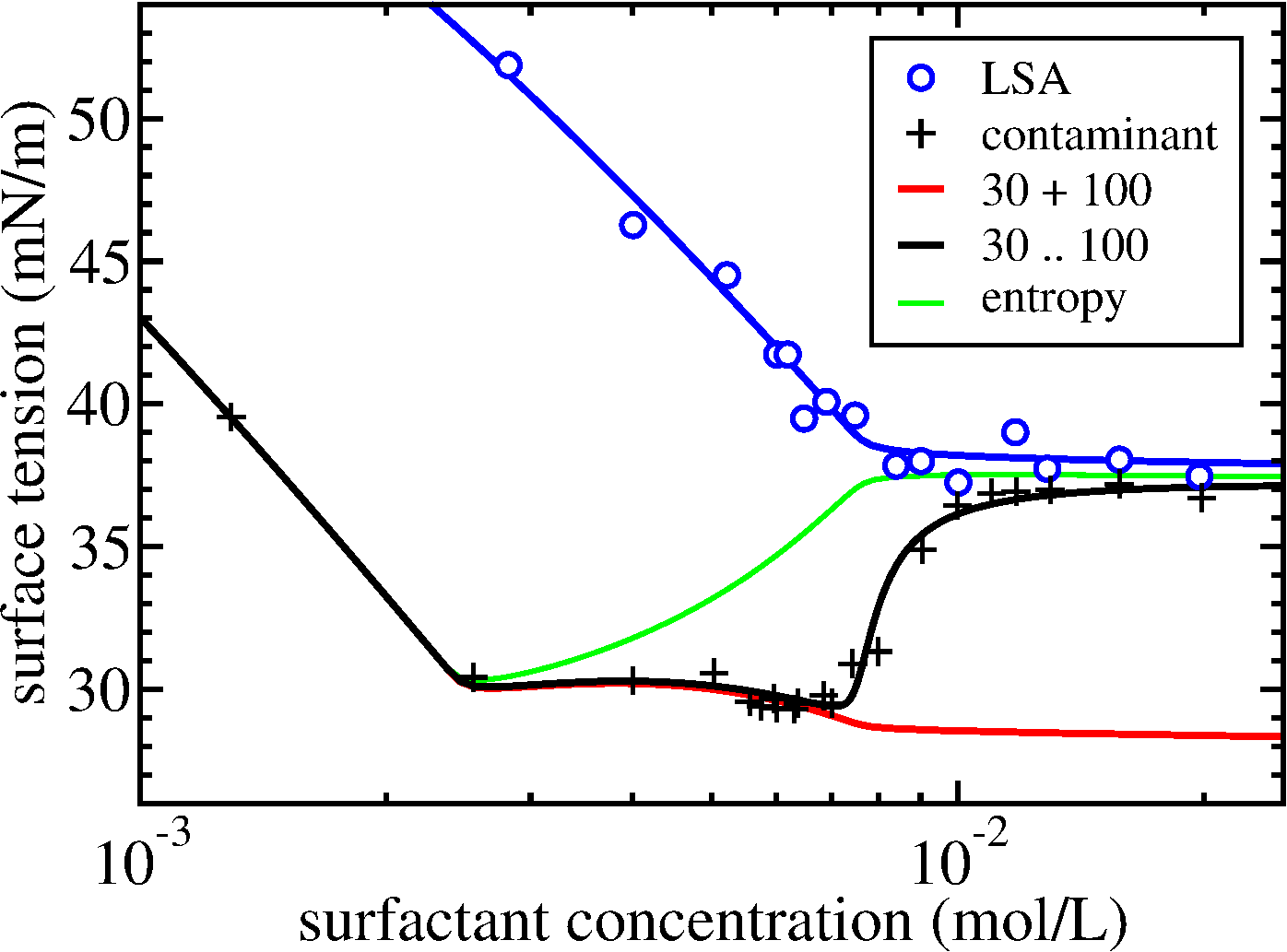}
\label{Fig:LSA_mixture_micel}}
\caption{Surface tension as a function of surfactant concentration for an aqueous solution of (a) C$_{10}$E$_8$ and (b) LSA. In (a), symbols are the experimental results by Lin {\em et al.} \cite{Lin_1999} for the purified C$_{10}$E$_8$ system (open symbols) and with 0.10\% dodecanol added (plus signs). In (b), symbols are the experimental results by McBain {\em et al.} \cite{McBain_1940a} for the purified LSA system (open symbols) and those containing an unknown amount of contaminant (plus signs). The solid blue lines are the theoretical results for the purified surfactant system. The other solid lines are the theoretical results in Section \ref{surfactant-contaminant_mixture_micelle_formation} for three different assumptions on the micellar composition: (1) solid red line: regular micelles + mixed micelles; (2) solid black line: range of (mixed) micelle compositions; (3) solid green line: no restriction on the composition range but with mixing entropy included. Values of the fit parameters are listed in Tables \ref{Table:data_single} and \ref{Table:data_mixture}.}
\label{Fig:Figure_3}
\end{figure}

Next, these formulas are used to describe surface tension experiments for three different surfactant solutions:\\
- purified C$_{12}$E$_8$ with 0.10\% dodecanol added \cite{Lin_1999} (Figure~\ref{Fig:C12E8_mixture_micel}),\\
- purified C$_{10}$E$_8$ with 0.10\% dodecanol added \cite{Lin_1999} (Figure~\ref{Fig:C10E8_mixture_micel}),\\
- Lauryl Sulfonic Acid (LSA) that is contaminated over time \cite{McBain_1940a} (Figure~\ref{Fig:LSA_mixture_micel}).

%
%

\vskip 5pt
\noindent
In all three experiments, we observe an abrupt deviation from dilute behaviour at a certain concentration that is well before, and clearly distinct from, the critical micelle concentration of the single surfactant system. We shall see that this {\em critical pre-micelle concentration} (cpc) signals the formation of {\em mixed} micelles containing species $b$ in a ratio ($m_b / m_a$) much higher than the bulk concentration ratio $c_b / c_a = \alpha$. 

To describe the experiments beyond the cpc, we determine, as a first step, the parameters $\Gamma_{a,\infty} \!=\! \Gamma_{\infty}$ and $K_a \!=\! K$ from the single surfactant system at low concentrations, as in Figure~\ref{Fig:C12E8_single}. Secondly, the micellar parameters $m$ and $x_0^a \!=\! x_0$ are determined from the single surfactant system at high concentrations, as in Figure~\ref{Fig:C12E8_single_micel}. Lastly, $\beta$ and $K_b$ are determined from the mixture at low concentrations, as in Figure~\ref{Fig:C12E8_mixture}. This leaves $x_0^b$ as the remaining fit parameter (essentially determined by the location of the cpc) for a given assumption on the micellar composition. Only in the case of LSA is $\alpha$ also unknown and to be determined by the fit. Below, we consider three different models for the micellar composition. 

\newpage
\noindent
{\bf i) Micellar composition: two types of micelle}
\vskip 5pt
\noindent
If one were to allow the presence of only one type of mixed micelle with a certain composition $m_b / m_a > \alpha$, the system would quickly run out of species $b$ necessary to form micelles at concentrations above the cmc \cite{Rusanov_book_1993}. It is therefore necessary to consider a mixture of at least {\bf two types of micelle}. As a minimum model, we take these two micelle types to be:\\
(1) regular micelles consisting only of species $a$ ($m_a \!=\! m$, $m_b \!=$ 0),\\
(2) mixed micelles with a {\em single value} $m_a \neq m$ and with $m_b \!=\! \beta \, (m - m_a) $.
\vskip 5pt
\noindent
The fixed value $m_a$ is a fit parameter.

The consequences of this assumption on the composition are shown as the solid red curves in Figures~\ref{Fig:C12E8_mixture_micel}, \ref{Fig:C10E8_mixture_micel} and \ref{Fig:LSA_mixture_micel}. All the red curves provide an excellent fit to the surface tension data for concentrations above the cpc but {\it only below the cmc}. The agreement is especially remarkable given the distinctly different behaviour of the surface tension in this intermediate concentration regime comparing the experimental results for C$_{12}$E$_8$ and C$_{10}$E$_8$ in Figures \ref{Fig:C12E8_mixture_micel} and \ref{Fig:C10E8_mixture_micel} to LSA in Figure \ref{Fig:LSA_mixture_micel}. Furthermore, the concentration region between the cpc and cmc is quite large for the LSA system and shows intricate, non-monotonous behaviour that is surprisingly well captured by the theoretical curve.

The red curves also show that {\bf above the cmc}, the assumption of only two type of micelles being present is no longer accurate (especially for LSA). It is seen that the experimental surface tension increases beyond the cmc to reach a plateau value, whereas the assumption of only two type of micelles leads to a leveling off of the surface tension directly at the cmc. This is an indication that in the experiments the composition of the mixed micelles (still) evolves beyond the cmc. To accommodate for this, we consider next a {\em range} in micellar composition.
\vskip 5pt
\noindent
{\bf ii) Micellar composition: composition range}
\vskip 5pt
\noindent
Next, we consider the situation where all values of the composition variable $m_a$ are allowed between a {\it minimum composition value} $m_a \!=\! m_{\rm min}$ up to $m_a \!=\! m$. The minimum composition value $m_{\rm min}$ is a fit parameter. Given that the red solid curves describe the experimental date well up to the cmc, the minimum value is expected to be close to the fixed value for $m_a$ of the previous composition model.

The consequences of allowing a composition range are shown as the black solid curves in Figures~\ref{Fig:C12E8_mixture_micel}, \ref{Fig:C10E8_mixture_micel} and \ref{Fig:LSA_mixture_micel}. It is observed that the agreement is drastically improved beyond the cmc. The model captures the experimental observation of an increase in surface tension beyond the cmc to reach a plateau value that is just below the value reached in the corresponding single surfactant system. The agreement is especially striking for LSA in Figure~\ref{Fig:LSA_mixture_micel}. This may, however, be somewhat fortuitous (even misleading) since in this case $\alpha$ has to be treated as an additional fit parameter and, furthermore, a precise value for $\beta$ could not really be determined due the lack of surface tension data at low concentrations.

Even though a clear improvement is observed when one considers a range in the micellar composition, the experimental data in Figures~\ref{Fig:C12E8_mixture_micel} and \ref{Fig:C10E8_mixture_micel} also show some shortcomings: the rise in surface tension beyond the cmc is less steep in the experiments and the plateau eventually reached in the experiments seems to be somewhat lower. Another short-coming of the model is the rather ad hoc introduction of a minimum value $m_{\rm min}$. Micelles containing less surfactant molecules of species $a$ are simply not allowed and one wonders whether this restriction can be lifted in a more natural way. An obvious candidate is to consider the gain in free energy associated with the {\em mixing entropy} of the two types of surfactant in the micelle. This is investigated next.
\vskip 5pt
\noindent
{\bf iii) Micellar composition: mixing entropy}
\vskip 5pt
\noindent
To determine the free energy of mixing, we count the number of ways $m$ positions on the surface of the micelle can be filled by $m_a$ surfactants:
\begin{equation}
W_{\rm mix} = \frac{1}{m} \, \frac{m!}{m_a! \, (m-m_a)!} \,,
\end{equation}
where $m_a$ runs from 1 to $m-1$. The factor $1/m$ in this expression accounts for the observation that the situation $m_a \!=\! 1$ (or $m_a \!=\! m - 1$) should correspond to only a single state $W_{\rm mix} \!=\! 1$ due to rotational symmetry.

\begin{figure}[ht]
\centering
\subfloat[monomer-micelle distribution]{\includegraphics[width=0.48\textwidth]{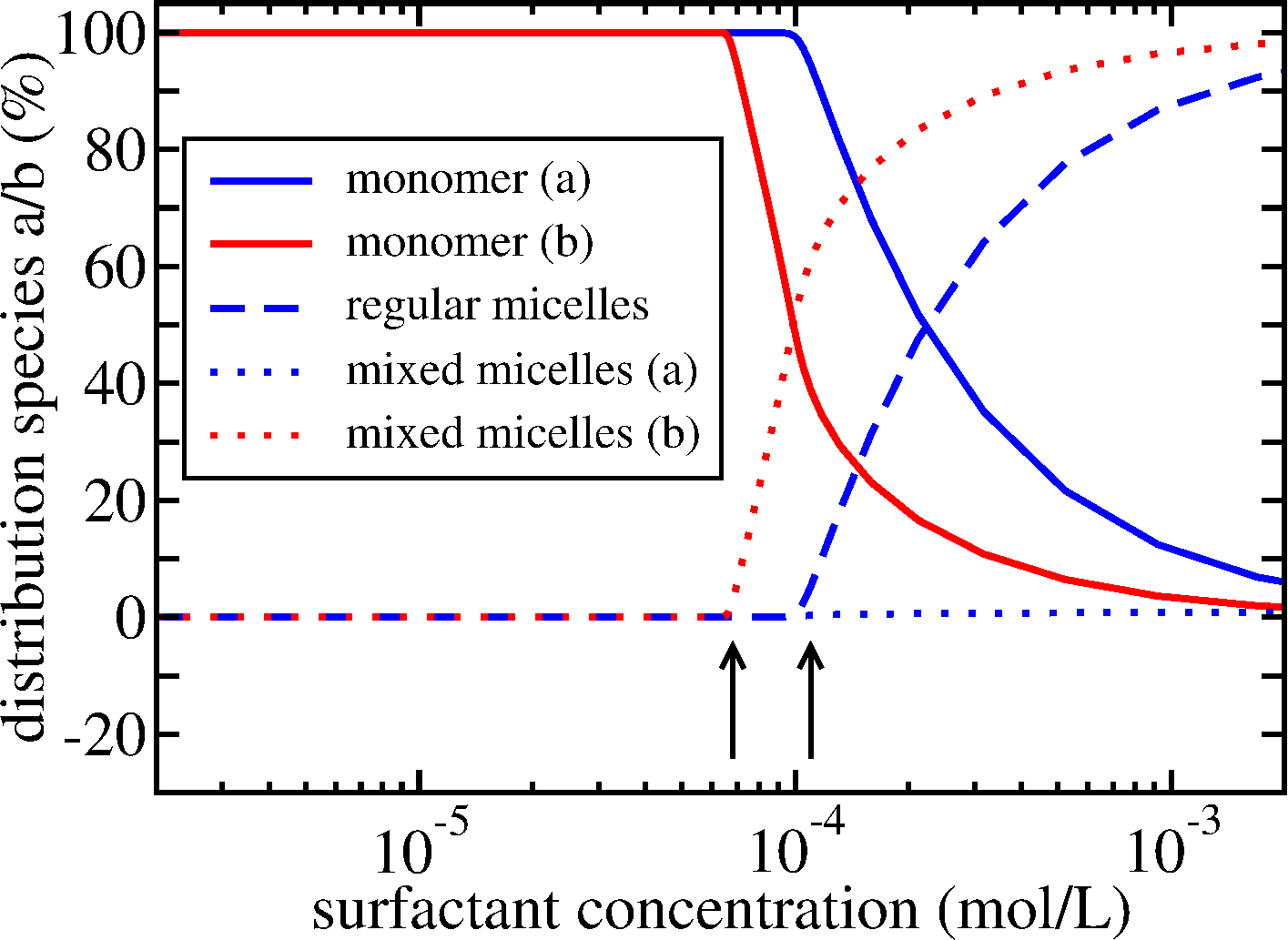}
\label{Fig:species_distribution}}
\hfill
\subfloat[micellar size distribution]{\includegraphics[width=0.48\textwidth]{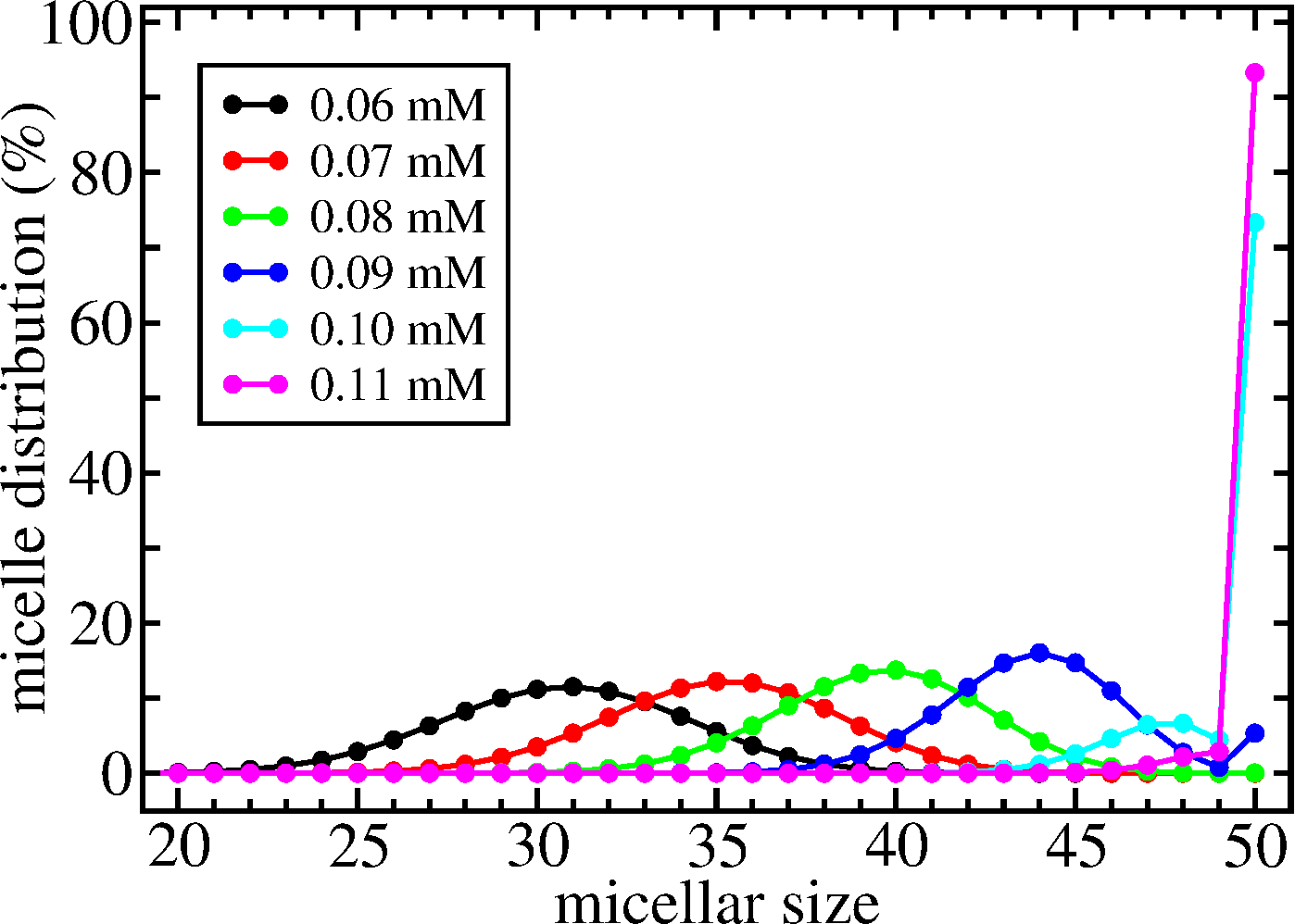}
\label{Fig:micellar_distribution}}
\caption{Calculated distributions for the C$_{12}$E$_8$ + contaminant system with mixing entropy included. This example corresponds to the solid green curve in Figure~\ref{Fig:C12E8_mixture_micel}. In (a), it is shown how the surfactant molecules (blue curves) and contaminant molecules (red curves) are distributed as monomers or as part of mixed or regular micelles as a function of surfactant concentration. The arrows indicate the approximate locations of the cpc $\approx$ 0.068 mM (left arrow) and cmc $\approx$ 0.11 mM (right arrow). In (b), the distribution of micelles is shown as a function of micellar size ($m_a$) for a number of surfactant concentrations.}
\label{Fig:Figure_4}
\end{figure}

Taking mixing entropy into account leads to the following adaptation of the micellar composition distribution in Eq.(\ref{eq:composition_distribution})
\begin{equation}
\frac{x_m^a}{m_a} = \left( \frac{x_1^a}{x_0^a} \right)^{\!\!m_a} \left( \frac{x_1^b}{x_0^b} \right)^{\!\!m_b} \, W_{\rm mix} \,.
\end{equation}
The inclusion of mixing entropy in the free energy lifts the restriction of a minimum value for $m_a$ in a natural way. A typical example of the distributions of the different species and their evolution as a function of concentration is shown in Figure~\ref{Fig:Figure_4}. A gradual shift in the average micellar size is observed which continues also above the cmc approaching the situation in which (almost) only regular micelles remain.

%
%

The effect on the surface tension of including surfactant-contaminant mixing entropy within the micelle mixing is shown as the solid green curves in Figures~\ref{Fig:C12E8_mixture_micel}, \ref{Fig:C10E8_mixture_micel} and \ref{Fig:LSA_mixture_micel}. For C$_{12}$E$_8$ and C$_{12}$E$_{10}$, the main feature is that the ultimate plateau value of the surface tension is reduced and is closer to the experimental value. For the LSA system, however, the inclusion of mixing entropy does not lead to such an improvement. This could be an indication that LSA and contaminant are not well mixed in the micelle, but such a conclusion cannot be made with certainty.

%
%

\section{Ionic surfactants}

\noindent
We have seen that the model introduced is able to capture the essence of the physical mechanisms involved in a quantitative manner for non-ionic surfactants. In this section, we aim to come to a similar description for {\bf ionic surfactants}. We have seen that for non-ionic surfactants, it is important to limit the number of parameters to guard against over-fitting \cite{Nguyen_2018}, but this adage holds even more for ionic surfactants. This means that we shall discard (notable) refinements made in the description of the single ionic surfactant system in order to limit the number of fit parameters. 

Again, before considering surfactant mixtures, we first discuss the situation of a single, ionic surfactant system.

\subsection{Single ionic surfactant type}

\noindent
We consider an aqueous solution of an ionic surfactant with concentration $c_{\rm s}$ to which salt may be added with a concentration $c_{\rm salt}$. We shall assume that the surfactant counterions are of the same type as the salt ions and that both surfactant and salt are strong electrolytes, fully dissociated in solution. For notational convenience, we assume that the solution consists of Sodium Dodecyl Sulfate (SDS) surfactant molecules and that the added salt is NaCl keeping in mind that the analysis applies more generally. Three type of ions (DS$^-$, Na$^+$, Cl$^-$) are then present in solution with respective (number) concentrations: 
\begin{equation}
c_{\rm DS} = c_{\rm s} \,, \hspace*{20pt} c_{\rm Na} = c_{\rm s} + c_{\rm salt} \,, \hspace*{20pt} c_{\rm Cl} = c_{\rm salt} \,.
\end{equation}
As before, the surfactant in solution is either present as a monomer or part of a micelle:
\begin{equation}
c_{\rm DS} = c_{\rm 1, DS} + c_{\rm m} \,.
\end{equation}
Also the counterions in solution are either free or part of the micelle. To accurately describe the micellar composition, it is necessary to introduce an additional fit parameter $r$ that denotes the {\em fraction of counterions that are part of a micelle} \cite{Mysels_1966}. This gives for the bulk counterion concentration
\begin{equation}
c_{\rm Na} = c_{\rm 1, Na} + r \, c_{\rm m} \,.
\end{equation}
The condition $r \!=$ 0 thus corresponds to the situation where all ionic surfactants in the micelle are dissociated, whereas $r \!=$ 1 corresponds to the situation where none of them are dissociated (effectively neutral in the micelle).

The third ion type present in solution, Cl$^-$, is always free in solution:
\begin{equation}
c_{\rm Cl} = c_{\rm 1, Cl} \,.
\end{equation}
Again, we shall assume that the chemical potential of the free ions is that of an infinitely dilute solution:
\begin{equation}
\label{eq:chemical_potential_ions}
\mu_{\rm i} = \mu^{\circ}_{\rm i} + k_{\rm B} T \, \ln(c_{1, i} / c^{\circ}) \,.
\end{equation}
where the index $i \!=$ DS$^-$, Na$^+$, Cl$^-$.

It can be argued that for experimental ionic systems deviations from this expression may be significant, even for a dilute system, when the concentration of added salt is large. Under such circumstances the expression used for the chemical potential of ion $i$ in solution with concentration $c_i$ is usually take to be of the following form
\begin{equation}
\label{eq:chemical_potential_gamma}
\mu_{\rm i} = \mu^{\circ}_{\rm i} + k_{\rm B} T \, \ln(\gamma_{\pm} \, c_i / c^{\circ}) \,.
\end{equation}
The activity coefficient $\gamma_{\pm}$ denotes the deviation from ideality and is then usually given by the following semi-empirical formula derived from Debye-H\"{u}ckel theory \cite{DH}:
\begin{equation}
^{10}\log(\gamma_{\pm}) = 0.055 \, I - \frac{0.5115 \, \sqrt{I}}{1 + 1.316 \, \sqrt{I}} \,,
\end{equation}
with the (dimensionless) ionic strength defined as $I \!=\! \frac{1}{2} \sum [c_i]$.

For the surfactant concentrations considered here, the factor $\gamma_{\pm}$ is close to unity in the absence of added salt. Furthermore, even when the added salt concentration is significant, the factor $\gamma_{\pm}$ is more or less constant for the range of surfactant concentrations considered \cite{Lucassen_1966, Prosser_2001}. The result is that we can disregard the factor $\gamma_{\pm}$ in Eq.(\ref{eq:chemical_potential_gamma}) and use Eq.(\ref{eq:chemical_potential_ions}) instead, but that we may then expect some salinity dependence of the fit parameters $K$ and $x_0$ at high concentrations of added salt.

As before, we use the Langmuir model to derive an expression for the surface tension. It is then necessary to make an assumption on the adsorption of counterions at the liquid-vapour surface. Here we shall consider the counterions to be {\bf fully bound} to the surface rendering the ionic surfactants essentially electrostatically neutral. The consequence is that the adsorption of counterions is equal to the adsorption of surfactant ions, $\Gamma_{\rm Na^+} \!=\! \Gamma_{\rm DS^-} \!\equiv \Gamma$, and that electrostatic contributions to the surface tension can be neglected.

Since we are mainly interested in the behaviour of the surface tension near the cmc and since it is important to limit the number of fit parameters, these approximations serve the purpose of the present article. It is, however, important to recognize the important contributions made to the theoretical description of the surface tension and adsorption of (mixtures of) ionic surfactants often leading to excellent agreement for different types of experiment \cite{Rosen_book, Davies_1951, Ingram_1980, Lucassen_1966, Lucassen_1981, Borwankar_1988, Blankschtein_1999, Mulqueen_1999, Vis_2015a, Vis_2015b} (see also the recent review in ref.\cite{Nguyen_2020} and references therein). In particular, we mention the work by Kralchevsky and coworkers \cite{Kralchevsky_1999, Kralchevsky_2002, Kralchevsky_2003, Kralchevsky_2014a, Kralchevsky_2014b} who included electrostatic and non-electrostatic interactions between adsorbed surfactant molecules and explicitly considered counterion binding in terms of an equilibrium constant $K_{\rm Stern}$ \cite{Kralchevsky_2014a, Kralchevsky_2014b}.

When the ionic surfactant is fully associated at the surface, the Langmuir model leads to the following expression for the surface tension (see the Supporting Information)
\begin{equation}
\label{eq:sigma_ions_x}
\sigma = \sigma_0 - k_{\rm B} T \,\, \Gamma_{\infty} \, \ln(1 + x_{\rm DS} \, x_{\rm Na}) \,,
\end{equation}
where $x_i$ ($i \!=$ DS$^-$, Na$^+$) is defined as
\begin{equation}
x_i \equiv \exp \, [ \, (\mu_{\rm i} - \mu^{\circ}_{\rm i} - \Delta E_{\rm s, i}) / k_{\rm B} T \, ] \,,
\end{equation}
and where $\Delta E_{\rm s, i}$ is the adsorption energy associated with the adsorption of ion $i$ from a (reference) bulk solution.

The surfactant and counterion adsorption can then be determined from Eq.(\ref{eq:sigma_ions_x}) by differentiation with respect to the respective chemical potential
\begin{equation}
\label{eq:Gamma_ions_x}
\Gamma_{\rm DS^-} = \Gamma_{\rm Na^+} \equiv \Gamma = \Gamma_{\infty} \, \frac{x_{\rm DS} \, x_{\rm Na}}{1 + x_{\rm DS} \, x_{\rm Na}} \,.
\end{equation}

Again, it is convenient to relate $x_i$ to the ion monomer concentration $c_{1,i} $. Inserting the expression for the chemical potential in Eq.(\ref{eq:chemical_potential_ions}) into the definition for $x_i$ gives:
\begin{equation}
\label{eq:sigma_ions}
\sigma = \sigma_0 - k_{\rm B} T \,\, \Gamma_{\infty} \, \ln(1 + K_2 \, c_{\rm 1, DS} \, c_{\rm 1, Na} \, ) \,,
\end{equation}
with
\begin{equation}
\label{eq:K_2}
K_2 = (v_0)^2 \, \exp \, [ \, - (\Delta E_{\rm s, DS} +  \Delta E_{\rm s, Na}) / k_{\rm B} T \, ] \,.
\end{equation}

\subsubsection{Single ionic surfactant -- dilute regime}

\begin{figure}[ht]
\centering
\subfloat[dilute regime]{\includegraphics[width=0.49\textwidth]{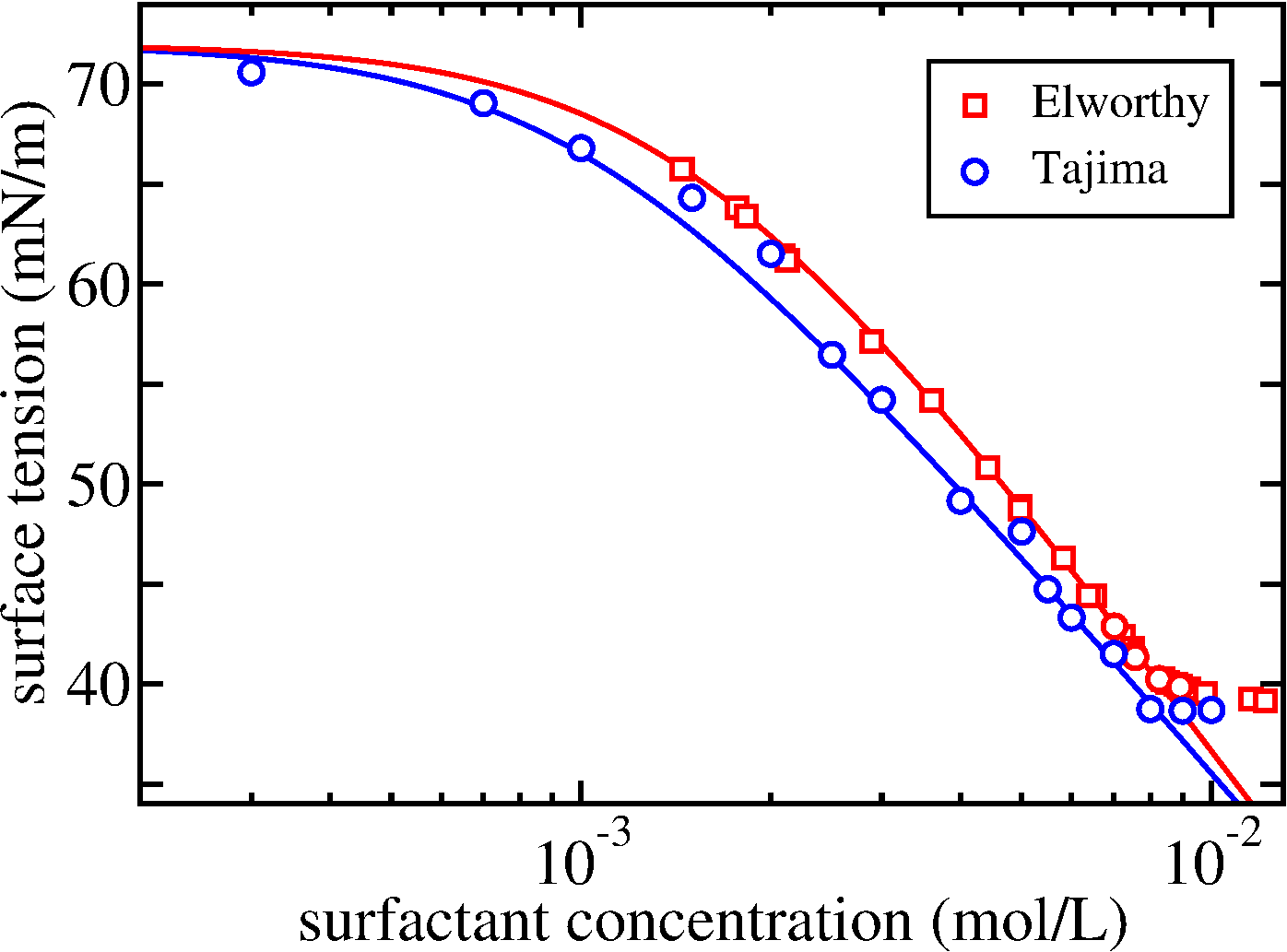}
\label{Fig:SDS_single}}
\hfill
\subfloat[micelle formation]{\includegraphics[width=0.49\textwidth]{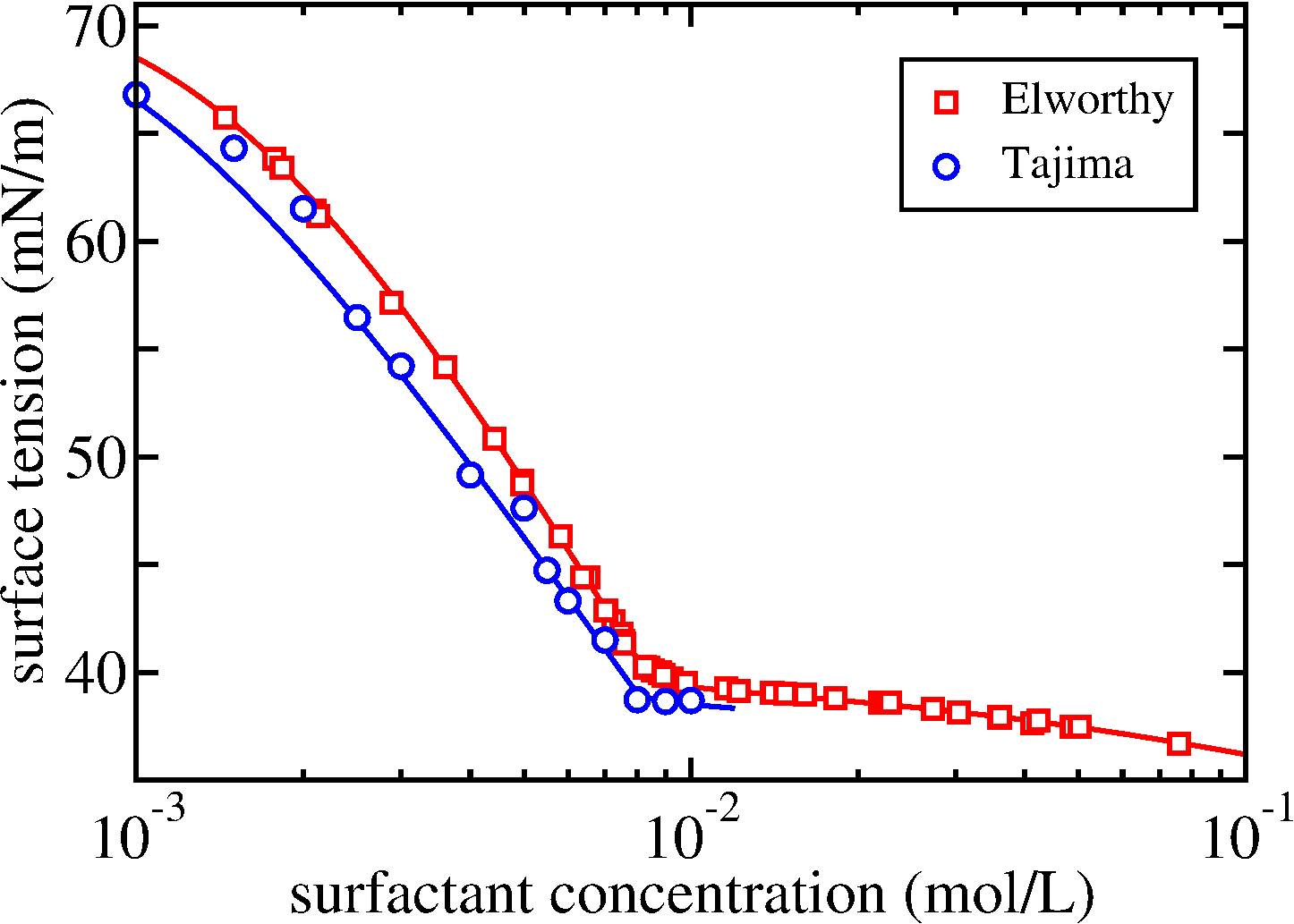}
\label{Fig:SDS_single_micelle}}
\caption{Surface tension as a function of surfactant concentration for an aqueous solution of purified SDS without added salt ($c_{\rm salt}\!=$ 0). Open squares are experimental results by Elworthy and Mysels \cite{Mysels_1966}. Open circles are experimental results by Tajima {\em et al.} \cite{Tajima_Sasaki_1970}. In (a), the solid lines are the Langmuir-Szyskowski-like equation in Eq.(\ref{eq:sigma_ions_LS}) with values for the two fit parameters $\Gamma_{\infty}$ and $K_2$ listed in Table \ref{Table:data_ions}. In (b), the solid lines are Eq.(\ref{eq:sigma_ions}) with values for the three additional fit parameters $r$, $m$ and $x_0$ from the mass action model listed in Table \ref{Table:data_ions}.}
\label{Fig:Figure_5}
\end{figure}

\noindent
Before considering micelle formation, we investigate the expression for $\sigma$ in Eq.(\ref{eq:sigma_ions}) in the {\em dilute regime}, where $c_{\rm m} \!\approx$ 0. The surface tension is then given by
\begin{equation}
\label{eq:sigma_ions_LS}
\sigma \approx \sigma_0 - k_{\rm B} T \,\, \Gamma_{\infty} \, \ln(1 + K_2 \, c_{\rm s} \, (c_{\rm s} + c_{\rm salt}) \, ) \hspace*{50pt} {\rm (dilute)}
\end{equation}
with the adsorption given by
\begin{equation}
\frac{\Gamma}{\Gamma_{\infty}} \approx \frac{K_2 \, c_{\rm s} \, (c_{\rm s} + c_{\rm salt})}{1 + K_2 \, c_{\rm s} \, (c_{\rm s} + c_{\rm salt}) \, } \hspace*{50pt} {\rm (dilute)}
\end{equation}
In Figure~\ref{Fig:SDS_single}, we compare the expression for the surface tension in Eq.(\ref{eq:sigma_ions_LS}) to experimental results for SDS by Elworthy and Mysels \cite{Mysels_1966} and by Tajima, Muramatsu, and Sasaki \cite{Tajima_Sasaki_1970}.

%
%

Again, up to a certain concentration, the surface tension is well described in terms of the surface tension $\sigma_0$ of the water-air interface, the slope $\Gamma_{\infty}$ and the cross-over concentration that is related to (the inverse square root of) $K_2$.

%
%

\begin{table}
\centering
\begin{tabular}{c || c | c || c | c | c}
SDS reference & $\Gamma_{\infty}$   & $1 / K_2$             & $r$ & $m$ & $x_0$ \\
              & 10$^{-6}$ mol/m$^2$ & 10$^{-6}$ (mol/L)$^2$ &     &     & \hspace*{1pt} 10$^{-7}$ \\
\hline
\hline
Elworthy {\em et al.} \cite{Mysels_1966}        & 3.70                           & 2.16 & 0.75 \cite{Mysels_1966} & 65 \cite{Mysels_1966} & 2.7 \\
Tajima {\em et al.} \cite{Tajima_Sasaki_1970}   & 3.19 \cite{Tajima_Sasaki_1970} & 1.00 & 0.75 \cite{Mysels_1966} & 65 \cite{Mysels_1966} & 2.5 \\
\end{tabular}
\caption{Values of the fit parameters used to plot the theoretical curves in Figures~\ref{Fig:Figure_5} and \ref{Fig:SDS_single_salt} for the aqueous solution of purified SDS. The first two columns are the fit parameters $\Gamma_{\infty}$ and $K_2$ from a fit of the data in the dilute concentration regime. The third and fourth and column are the micellar fit parameters $r$ and $m$ taken from ref.\cite{Mysels_1966}. The final column is the value of the micellar fit parameter $x_0$.}
\label{Table:data_ions}
\end{table}

\subsubsection{Single ionic surfactant type -- micelle formation}

\noindent
When micelles form, the surface tension is no longer described by the approximation in Eq.(\ref{eq:sigma_ions_LS}). In order to use the full expression for the surface tension in Eq.(\ref{eq:sigma_ions}) instead, we need to model micelle formation. We shall assume that each micelle is composed of $m$ DS$^-$ surfactant molecules with $r \, m$ Na$^+$ counterions adsorbed. Following the original analysis by Elworthy {\em et al.} \cite{Mysels_1966}, we take the fraction $r \!=$ 0.75 and set the micellar size $m \!=$ 65. We further introduce $\Delta E_{\rm m, DS}$ and $\Delta E_{\rm m, Na}$ as the energy gain for a DS$^-$ or Na$^+$ ion to become part of the micelle, although only the combination $\Delta E_{\rm m, DS} + r \, \Delta E_{\rm m, Na}$ leads to an independent fit parameter. 

The free energy is then given by
\begin{eqnarray}
&& \frac{v_0}{V} \, F(x_{\rm 1, DS},x_{\rm 1, Na}, x_m) = x_{\rm 1, DS} \, k_{\rm B} T \, (\ln(x_{\rm 1, DS}) - 1 )
+ x_{\rm 1, Na} \, k_{\rm B} T \, (\ln(x_{\rm 1, Na}) - 1 ) \nonumber \\
&& \hspace*{20pt} + \frac{x_m}{m} \, k_{\rm B} T \, (\ln(\frac{x_m}{m}) - 1) + x_m \, (\Delta E_{\rm m, DS} + r \, \Delta E_{\rm m, Na}) \,.
\end{eqnarray}
The first three terms represent the translational entropy of surfactant monomers, free counterions and micelles. The last term represents the energy gain of surfactant and counterion to be part of the micelle. The free energy is to be minimized with respect to $x_{\rm 1, DS}$, $x_{\rm 1, Na}$ and $x_m$ under the constraint that $v_0 \, c_{\rm s} \!=\! X_{\rm s} \!=\! x_{\rm 1, DS} + x_m$ and $v_0 \, (c_{\rm s} + c_{\rm salt}) \!=\! X_{\rm s} + X_{\rm salt} \!=\! x_{\rm 1, Na} + r \, x_m$. The minimization leads to the following expression for the surfactant volume fraction $x_m$
\begin{equation}
\label{eq:composition_distribution_ions}
x_m = m \left( \frac{x_{\rm 1, DS} \, (x_{\rm 1, Na})^r}{x_0} \right)^{\!m} \,,
\end{equation}
where we have defined
\begin{equation}
x_0 \equiv \exp \, [ \, (\Delta E_{\rm m, DS} + r \, \Delta E_{\rm m, Na}) / k_{\rm B} T \, ] \,.
\end{equation}
We are now in a position to calculate the surface tension from the full expression in Eq.(\ref{eq:sigma_ions}) using Eq.(\ref{eq:composition_distribution_ions}) to determine $x_{\rm 1, DS} \!=\! v_0 \, c_{\rm 1, DS}$ and $x_{\rm 1, Na} \!=\! v_0 \, c_{\rm 1, Na}$. In Figure~\ref{Fig:SDS_single_micelle} we compare the result to the experimental results for SDS in refs.\cite{Mysels_1966, Tajima_Sasaki_1970}. With $r$ and $m$ taken from ref.\cite{Mysels_1966}, the value of the single, remaining fit parameter $x_0$ is determined by the location of the cmc. As already concluded by Elworthy and Mysels \cite{Mysels_1966}, satisfactory agreement is obtained for concentrations both below and above the cmc.

The agreement is especially striking considering the large number of simplifications made: (1) attractive, van der Waals-like interactions between the surfactant molecules adsorbed to the surface are not taken into account in the Langmuir model (as they are in the Frumkin model \cite{Frumkin} or other surface EOS models \cite{Rosen_book}), (2) the ionic surfactant is considered not to be dissociated at the surface \cite{Kralchevsky_2014a, Kralchevsky_2014b}, and (3) electrostatic contributions to the surface tension have been neglected \cite{Kralchevsky_1999, Lucassen_1966}.

%
%

\begin{figure}[ht]
\centering
\includegraphics[width=0.7\textwidth]{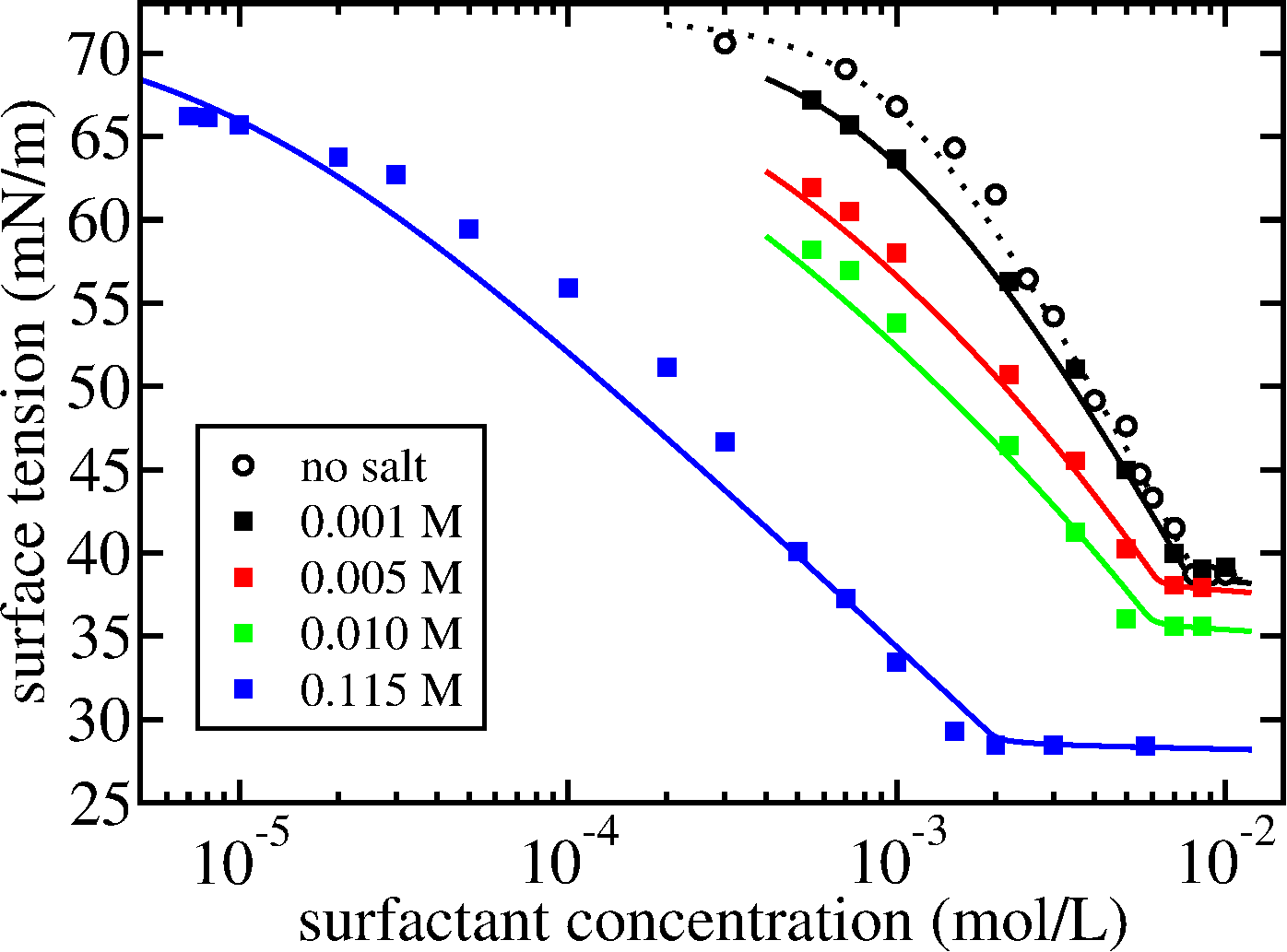}
\caption{Surface tension as a function of surfactant concentration for an aqueous solution of purified SDS for different concentrations of added salt. Open circles are the experimental results by Tajima {\em et al.} \cite{Tajima_Sasaki_1970} in Figure~\ref{Fig:SDS_single_micelle} with the corresponding theoretical fit as the dotted line. Closed square symbols are the experimental results for different concentrations of added salt in ref.\cite{Tajima_1970} ($c_{\rm salt}\!=$ 0.115 M) and ref.\cite{Tajima_1971} ($c_{\rm salt}\!=$ 0.001 M, 0.005 M, 0.01 M). The solid lines are calculated from Eq.(\ref{eq:sigma_ions}) without additional fitting using the same parameters in Table \ref{Table:data_ions} as for SDS results by Tajima {\em et al.} \cite{Tajima_Sasaki_1970}. Only the two highest salt concentrations show some salt dependence in the value of $x_0$, $x_0 \!=$ 3.1 $\times 10^{-7}$ ($c_{\rm salt}\!=$ 0.01 M) and $x_0 \!=$ 4.7 $\times 10^{-7}$ ($c_{\rm salt}\!=$ 0.115 M).}
\label{Fig:SDS_single_salt}
\end{figure}

As a further test of the assumption to consider the liquid surface essentially as electrostatically neutral, one may consider the influence of the amount of added salt on the surface tension. In Figure~\ref{Fig:SDS_single_salt} we compare the expression for the surface tension in Eq.(\ref{eq:sigma_ions}) with the experiments in refs.\cite{Tajima_1970} and \cite{Tajima_1971} for various salt concentrations up to $c_{\rm salt}\!=$ 0.115 M. For concentrations below the cmc, the surface tension is well approximated by Eq.(\ref{eq:sigma_ions_LS}) with the fit parameters $\Gamma_{\infty}$ and $K_2$ equal to those determined by the experiments in the {\em absence of added salt}. Even though some differences with the experimental results are present, overall trends are very well reproduced showing that there is no real need to introduce any salinity dependence of these fit parameters in this concentration regime \cite{Prosser_2001}. Eq.(\ref{eq:sigma_ions_LS}) also describes the experiments well at higher surfactant concentrations but then it is necessary to allow some salinity dependence in $x_0$, at the two highest added salt concentrations, to correctly match the location of the cmc.

\subsection{Ionic surfactant-contaminant mixture}

\noindent
We now extend the previous analysis for ionic surfactants to include the contaminant as an additional component. The three species then involved are denoted as species $a$ (ionic surfactant DS$^-$), species $b$ (contaminant DOH) and species $c$ (counterions Na$^+$), with respective concentrations $c_a \!=\! c_{\rm s}$, $c_b \!=\! \alpha \, c_{\rm s}$ and $c_c \!=\! c_{\rm s} + c_{\rm salt}$. 

In the context of the assumptions made previously, one can shown that the Langmuir model leads to the following expression for the surface tension of such a mixture
\begin{equation}
\label{eq:sigma_ions_mixture_x}
\sigma = \sigma_0 - k_{\rm B} T \,\, \Gamma_{a,\infty} \, \ln(x_a \, x_c + (1 + x_b)^{\beta}) \,,
\end{equation}
where, analogously to before, $x_i$ ($i = a, b, c$) is defined as 
\begin{equation}
x_i \equiv \exp \, [ \, (\mu_i - \mu^{\circ}_i - \Delta E_{\rm s, i}) / k_{\rm B} T \, ] \,,
\end{equation}
and where $\Delta E_{\rm s, i}$ is the adsorption energy associated with the adsorption of species $i$ from a (reference) bulk solution. Again, the parameter $\beta$ accounts for the fact that a surfactant molecule ($a$) may take up more of the available area than a contaminant molecule ($b$) due to a possible difference in size of the polar head group. The fully adsorbed counterions ($c$) are assumed not to reduce the available surface area.

The adsorption of surfactant ($a$) and contaminant ($b$) can be determined from Eq.(\ref{eq:sigma_ions_mixture_x}) by differentiation with respect to the respective chemical potential
\begin{equation}
\label{eq:Gamma_ions_mixture_x}
\frac{\Gamma_a}{\Gamma_{a,\infty}} = \frac{x_a \, x_c}{x_a \, x_c + (1 + x_b)^{\beta}} 
\hspace*{20pt} {\rm and} \hspace*{20pt}
\frac{\Gamma_b}{\Gamma_{b,\infty}} = \frac{x_b \, (1 + x_b)^{\beta-1}}{x_a \, x_c + (1 + x_b)^{\beta}} \,,
\end{equation}
where $\beta \!=\! \Gamma_{b,\infty} / \Gamma_{a,\infty} \geqslant$ 1.

Again, it is convenient to relate $x_i$ to the surfactant monomer concentration $c_{1,i}$. Inserting the expression for the chemical potentials in Eq.(\ref{eq:chemical_potential_ions}) into the definition for $x_i$ gives:
\begin{equation}
\label{eq:sigma_ions_mixture}
\boxedB{ \sigma = \sigma_0 - k_{\rm B} T \,\, \Gamma_{a,\infty} \, \ln(K_a \, c_{1, a} \, c_{1, c} + (1 + K_b \, c_{1, b})^{\beta}) }
\end{equation}
with
\begin{eqnarray}
\label{eq:K_a_and_K_b}
K_a &=& (v_0)^2 \, \exp \, [ \, - (\Delta E_{\rm s, DS} +  \Delta E_{\rm s, Na}) / k_{\rm B} T \, ] \,, \nonumber \\
K_b &=& v_0 \, \exp \, [ \, - \Delta E_{\rm s, DOH} / k_{\rm B} T \, ] \,.
\end{eqnarray}

Next, the expression for the surface tension in Eq.[\ref{eq:sigma_ions_mixture}] is used to describe two experiments on SDS solutions:\\
- purified SDS + 0.20\% dodecanol, by Vollhardt {\em et al.} \cite{Vollhardt_2000} (Figures \ref{Fig:SDS_mix_Vollhardt} and \ref{Fig:SDS_mix_Vollhardt_micelle}),\\
- contaminated SDS, by Razavi {\em et al.} \cite{Razavi_2022} (Figures \ref{Fig:SDS_mix_Razavi} and \ref{Fig:SDS_mix_Razavi_micelle}).
\vskip 5pt
\noindent
In both examples, values of the various SDS parameters are taken from the purified SDS experiments by Tajima {\em et al.} \cite{Tajima_Sasaki_1970} in Table \ref{Table:data_ions} to be consistent with the results of the uncontaminated surfactant system, i.e. $\Gamma_{a,\infty} \!=\! \Gamma_{\infty}$ and $K_a \!=\! K_2$.

\subsubsection{Ionic surfactant-contaminant mixture -- dilute regime}

\noindent
In the dilute regime, no micelles are present so that the concentration of monomers in solution for each species is equal to the total concentration of that species. In that case the expression for the surface tension in Eq.(\ref{eq:sigma_ions_mixture}) becomes
\begin{equation}
\label{eq:sigma_ions_mixture_LS}
\sigma \approx \sigma_0 - k_{\rm B} T \,\, \Gamma_{a,\infty} \, \ln(K_a \, c_{\rm s} \, (c_{\rm s} + c_{\rm salt}) + (1 + K_b \, \alpha \, c_{\rm s})^{\beta}) \hspace*{50pt} {\rm (dilute)}
\end{equation}
%

%
%

\begin{figure}[ht]
\centering
\subfloat[Vollhardt et al.]{\includegraphics[width=0.49\textwidth]{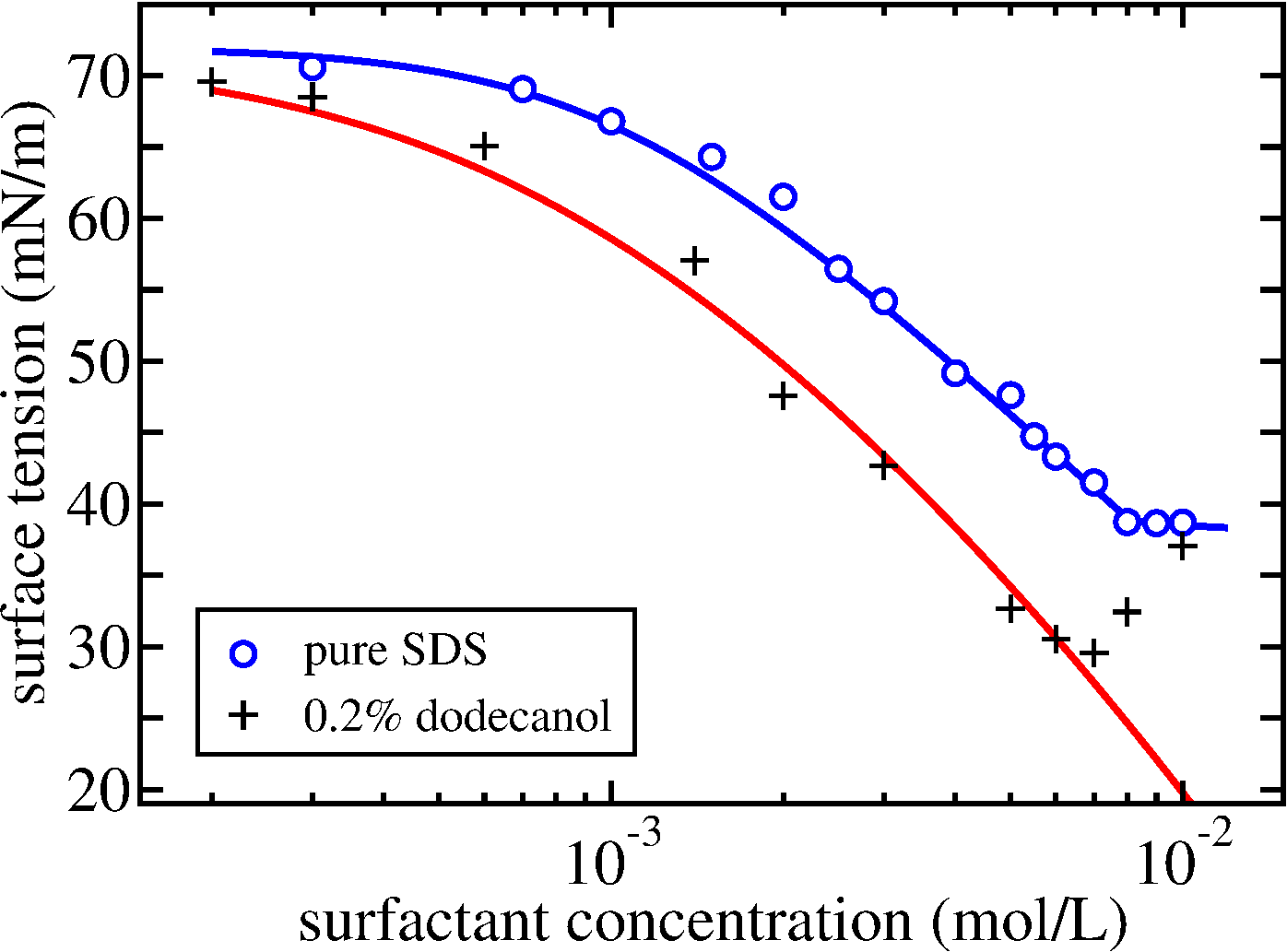}
\label{Fig:SDS_mix_Vollhardt}}
\hfill
\subfloat[Razavi et al.]{\includegraphics[width=0.49\textwidth]{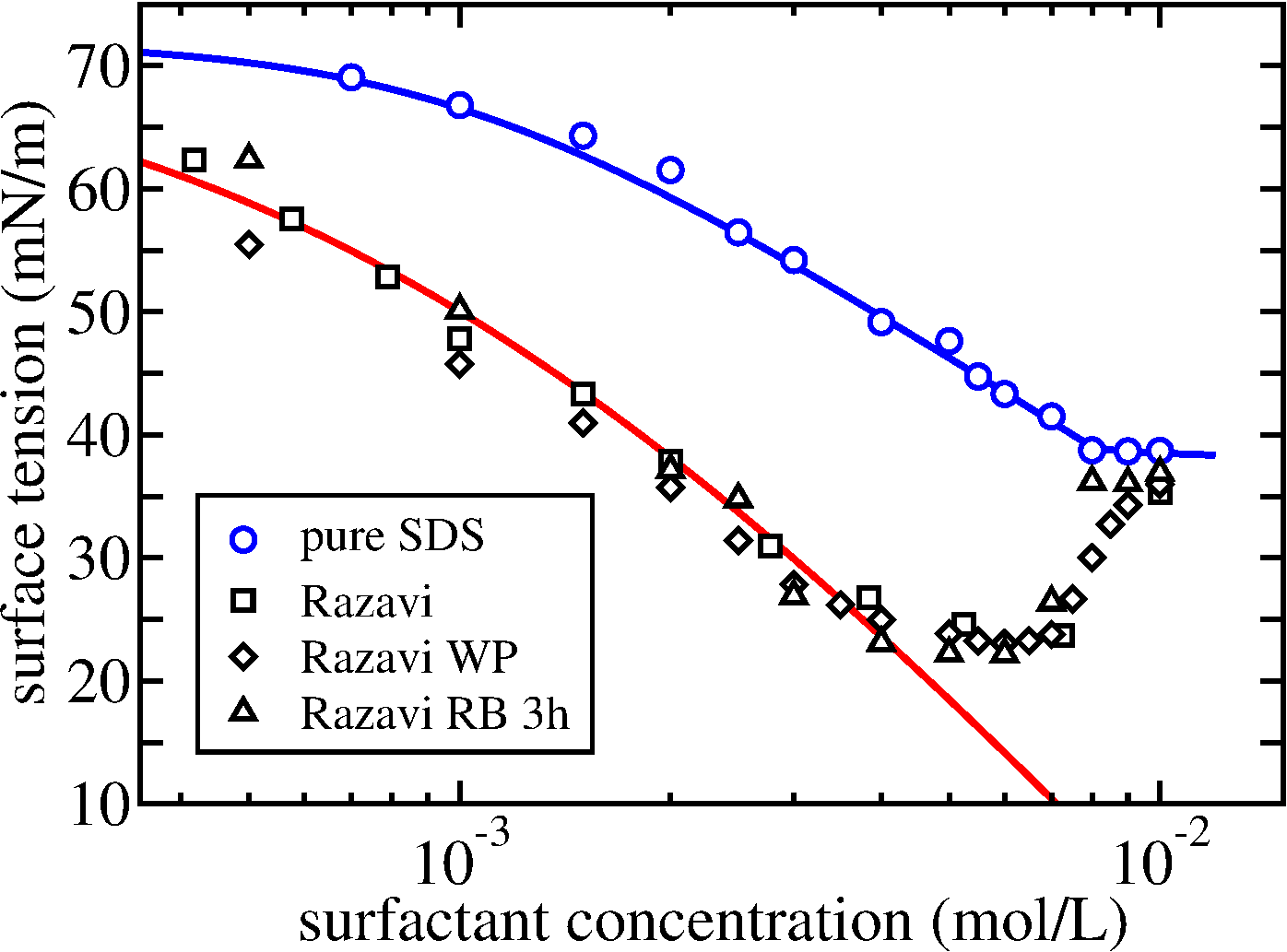}
\label{Fig:SDS_mix_Razavi}}
\caption{Surface tension as a function of surfactant concentration for an aqueous solution of SDS without added salt. Open circles are the experimental results by Tajima {\em et al.} \cite{Tajima_Sasaki_1970} for purified SDS shown in Figure~5 with the corresponding theoretical fit as the solid blue line. In (a), the plus signs are the experimental results by Vollhardt {\em et al.} \cite{Vollhardt_2000} for purified SDS with 0.20\% dodecanol added. In (b), the black symbols are the experimental results by Razavi {\em et al.} \cite{Razavi_2022} containing an unknown amount of contaminant for three different measurement procedures. The solid red lines are Eq.(\ref{eq:sigma_ions_mixture_LS}) with values for the additional fit parameters $\beta$ and $K_b$ listed in Table \ref{Table:data_ions_mixture}.}
\label{Fig:Figure_7}
\end{figure}

\noindent
In Figure \ref{Fig:Figure_7}, we show the experimental results by Vollhardt {\em et al.} \cite{Vollhardt_2000} and by Razavi {\em et al.} \cite{Razavi_2022}, respectively, together with Eq.(\ref{eq:sigma_ions_mixture_LS}) shown as the solid red lines. The two fit parameters $\beta$ and $K_b$ are determined from the experimental results in the dilute regime (see Table \ref{Table:data_ions_mixture}). The amount of contaminant $\alpha$ in the experiments by Razavi {\em et al.} \cite{Razavi_2022} is unknown and also needs to be fitted by the data. The fit value $\beta \!=$ 3.5 obtained is consistent with independent measurements of the adsorption of the fully saturated pure SDS system, $\Gamma_{a, \infty} \!=$ 3.19 $\times$ 10$^{-6}$ mol/m$^2$ (Table 1 of ref.\cite{Tajima_Sasaki_1970}), and of pure dodecanol on water, $\Gamma_{b, \infty} \!=$ 11.1 $\times$ 10$^{-6}$ mol/m$^2$ (Table 2 of ref.\cite{Lin_1999}), which leads to $\beta = \Gamma_{b, \infty} / \Gamma_{a, \infty} \!=$ 3.48. Figure \ref{Fig:Figure_7} shows that up to a certain concentration the experimental results are well reproduced by the solid red lines but that, at higher concentrations, micelle formation needs to be included.

\subsubsection{Ionic surfactant-contaminant mixture -- micelle formation}

\noindent
Again, to describe micelle formation, we first need to consider the composition of the micelles. We shall assume that a micelle consists of $m_a$ ions of the dominant species $a$ (the ionic surfactant DS$^-$) and $m_b$ molecules of species $b$ (the contaminant DOH). Furthermore, to each micelles $r \, m_a $ counterions (Na$^+$) are attached. Again, it is assumed that if the number of molecules of species $a$ in a micelle is less than $m$, all the available surface area of the micelle is then filled up by species $b$, i.e. $m_b \!=\! \beta \, (m - m_a)$. We shall further introduce 
$\Delta E_{\rm m, DS}$, $\Delta E_{\rm m, DOH}$, and $\Delta E_{\rm m, Na}$ as the energy gain of each species to become part of the micelle.

%
%

\begin{figure}[ht]
\centering
\subfloat[Vollhardt et al.]{\includegraphics[width=0.49\textwidth]{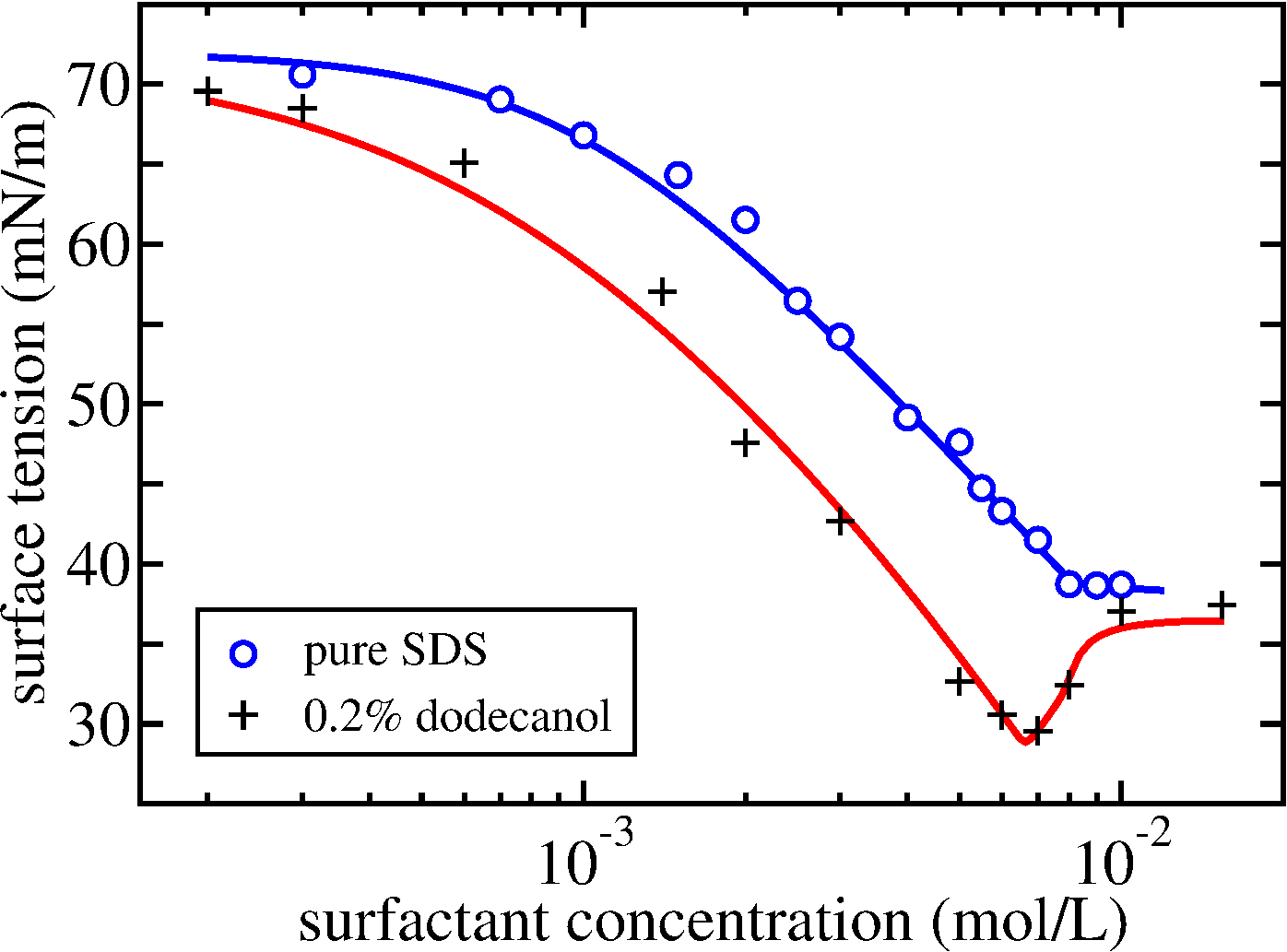}
\label{Fig:SDS_mix_Vollhardt_micelle}}
\hfill
\subfloat[Razavi et al.]{\includegraphics[width=0.49\textwidth]{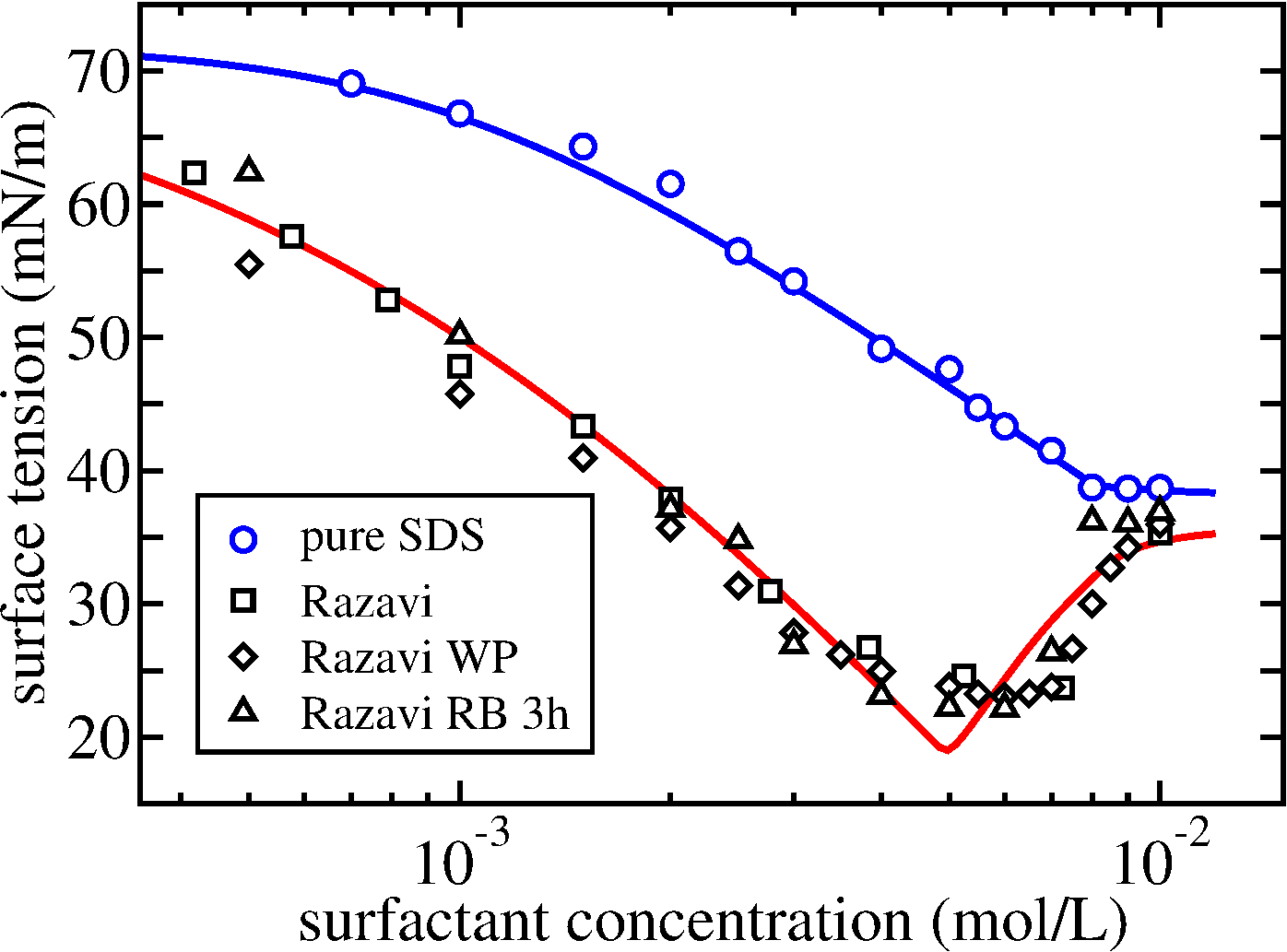}
\label{Fig:SDS_mix_Razavi_micelle}}
\caption{Surface tension as a function of surfactant concentration for an aqueous solution of SDS without added salt. Open circles are the experimental results by Tajima {\em et al.} \cite{Tajima_Sasaki_1970} for purified SDS shown in Figure~5 with the corresponding theoretical fit as the solid blue line. In (a), the plus signs are the experimental results by Vollhardt {\em et al.} \cite{Vollhardt_2000} for purified SDS with 0.20\% dodecanol added shown in Figure~\ref{Fig:SDS_mix_Vollhardt}. In (b), the black symbols are the experimental results by Razavi {\em et al.} \cite{Razavi_2022} shown in Figure~\ref{Fig:SDS_mix_Razavi}. Both solid red lines are Eq.(\ref{eq:sigma_ions_mixture}) with the same pre-micellar composition range and value of the fit parameter $x_0^{\rm b}$ listed in Table \ref{Table:data_ions_mixture}.}
\label{Fig:Figure_8}
\end{figure}

As the system may comprise micelles of different compositions, the micellar free energy contribution is again, in principal, a sum over all integer values between 0 and $m$ of the composition variable $m_a$. The free energy thus becomes a function of the three volume fractions $x_1^a$, $x_1^b$ and $x_1^c$ and the {\em distributions} $\{x_m^a\}$ and $\{x_m^b\}$
\begin{eqnarray}
&& \frac{v_0}{V} \, F(x_1^a,\{x_m^a\},x_1^b,\{x_m^b\},x_1^c) = \sum_{i} x_1^i \, k_{\rm B} T \, (\ln(x_1^i) - 1) \\
&& + \sum_{m_a} \left[ \frac{x_m^a}{m_a} \, k_{\rm B} T \, (\ln(\frac{x_m^a}{m_a}) - 1) + x_m^a \, (\Delta E_{\rm m, DS} + r \, \Delta E_{\rm m, Na}) + x_m^b \, \Delta E_{\rm m, DOH} \right] \nonumber \,.
\end{eqnarray}
The first term denotes the translational entropy of the free species ($i = a, b, c$). The second term comprises a summation over all values of $m_a$ of the free energy of micelles consisting of the micellar translational entropy and energy gain. The free energy is to be minimized with respect to $x_1^a$, $x_1^b$, $x_1^c$ and the distributions $\{x_m^a\}$ and $\{x_m^b\}$ under the constraint that $v_0 \, c_{\rm s} \!=\! X_{\rm s} \!=\! x_1^a + \sum x_m^a$, $\alpha X_s \!=\! x_1^b + \sum x_m^b$, $v_0 \, (c_{\rm s} + c_{\rm salt}) \!=\! X_{\rm s} + X_{\rm salt} \!=\! x_1^c + \sum x_m^b$ and $x_m^a / m_a \!=\! x_m^b / m_b$. The minimization leads to the following expression for the micellar composition distribution
\begin{equation}
\label{eq:composition_distribution_ions_mixture}
\frac{x_m^a}{m_a} = \left( \frac{x_1^a}{x_0^a} \right)^{\!\!m_a} \left( \frac{x_1^b}{x_0^b} \right)^{\!\!m_b} \left( x_1^c \right)^{r m_a} \,,
\end{equation}
where $m_a$ runs over the values allowed and where we have defined
\begin{eqnarray}
x_0^a &\equiv& \exp \, [ \, (\Delta E_{\rm m, DS} + r \, \Delta E_{\rm m, Na}) / k_{\rm B} T \, ]  \,, \nonumber \\
x_0^b &\equiv& \exp \, [ \, \Delta E_{\rm m, DOH} / k_{\rm B} T \, ]  \,.
\end{eqnarray}
%

%
%

\begin{table}
\centering
\begin{tabular}{c || c | c | c || c | c }
SDS + contaminant & $\alpha$ \hspace*{1pt} & \hspace*{5pt} $\beta$ \hspace*{5pt} & $1 / K_b$                                     & composition & \hspace*{1pt} $x_0^{\rm b}$ \\
reference         & ($\%$)                 &                                     & \hspace*{1pt} 10$^{-6}$ (mol/L) \hspace*{1pt} & interval    & \hspace*{1pt} 10$^{-7}$ \\
\hline
\hline
Vollhardt {\em et al.} \cite{Vollhardt_2000} \hspace*{1pt} & 0.20 \cite{Vollhardt_2000} & 3.5 & 3.75 & 50 .. 65 & 2.1 \\
Razavi {\em et al.} \cite{Razavi_2022} \hspace*{1pt}       & 0.44                       & 3.5 & 3.75 & 50 .. 65 & 2.1 \\
\end{tabular}
\caption{Values of the additional fit parameters used to plot the theoretical curves in Figures~\ref{Fig:Figure_7} and \ref{Fig:Figure_8} for the SDS with contaminant system. The first three columns are the fit parameters $\alpha$, $\beta$ and $K_b$ in the dilute concentration regime. The amount of dodecanol ($\alpha \!=$ 0.20\%) for Vollhardt {\em et al.} is set by the experimental conditions in ref.\cite{Vollhardt_2000}. The fourth column indicates the composition of the mixed micelles. The final column is the fit parameter $x_0^{\rm b}$ used to describe micelle formation.}
\label{Table:data_ions_mixture}
\end{table}

In Figure \ref{Fig:Figure_8}, the full expression for the surface tension in Eq.(\ref{eq:sigma_ions_mixture}), using Eq.(\ref{eq:composition_distribution_ions_mixture}) to determine the surfactant monomer concentrations $x_{\rm 1,i} \!=\! v_0 \, c_{\rm 1,i}$, is compared to the experimental results for SDS \cite{Vollhardt_2000, Razavi_2022}. The theoretical curves (solid red lines) are both determined using the parameters $\Gamma_{a,\infty} \!=\! \Gamma_{\infty}$ and $K_a \!=\! K_2$ from the single surfactant system at low concentrations, the micellar parameters $m$, $r$ and $x_0^a \!=\! x_0$ from the single surfactant system at high concentrations and $\alpha$, $\beta$ and $K_b$ determined from the mixture at low concentrations. This only leaves the pre-micellar composition range and $x_0^b$ as fit parameter to determine both theoretical curves in Figure \ref{Fig:Figure_8} (see Table \ref{Table:data_ions_mixture}). The agreement is striking especially given the fact that only the value of $\alpha$ differs between the two sets of experiments.

\section{Discussion and Conclusions}

\noindent
The aim in this article is to arrive at a quantitative description of the minimum in the surface tension observed in experiments on certain aqueous surfactant solutions containing a small amount of contaminant. To achieve this goal, a relatively simple approach is used based on the Langmuir model for adsorption in combination with the mass action model for micelle formation. Key in the theoretical description is an expression for the surface tension, Eq.(\ref{eq:sigma_mixture}) or Eq.(\ref{eq:sigma_ions_mixture}), derived from a Statistical Thermodynamic treatment of the Langmuir model extended to describe surfactant mixtures with different (molar) surface areas. Since the model is able to reproduce quite well experimental results for the minimum in both non-ionic and ionic surfactant systems, we can draw some conclusions on the physical picture that now emerges.

\begin{figure}[ht]
\centering
\subfloat[adsorption]{\includegraphics[width=0.49\textwidth]{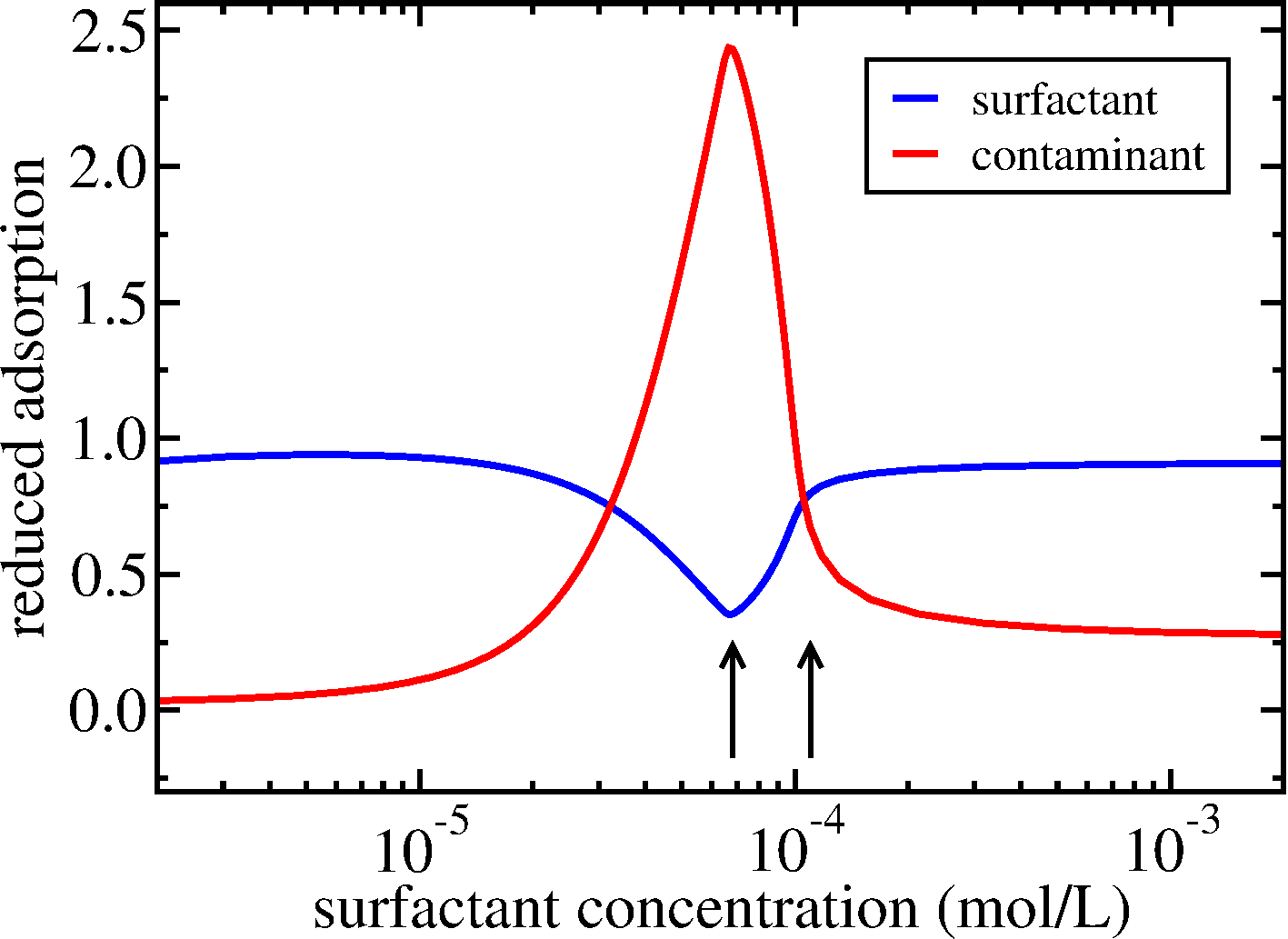}
\label{Fig:adsorption}}
\hfill
\subfloat[chemical potential]{\includegraphics[width=0.49\textwidth]{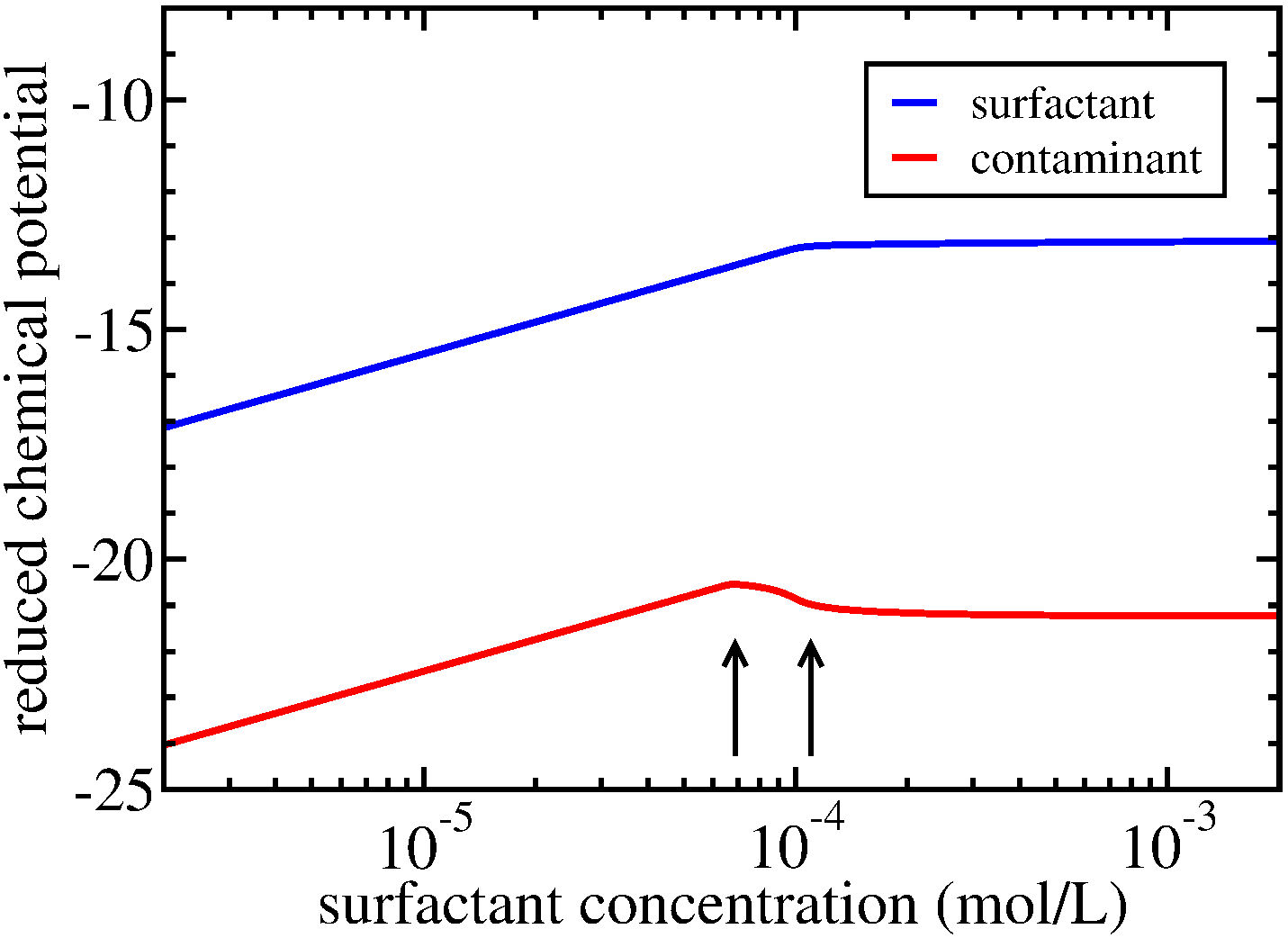}
\label{Fig:chemical_potential}}
\caption{Calculated adsorption and chemical potential for the C$_{12}$E$_8$ + contaminant system with mixing entropy included as a function of surfactant concentration. This example corresponds to the solid green curve in Figure~\ref{Fig:C12E8_mixture_micel}. In (a), the reduced adsorptions are defined as $\Gamma_a / \Gamma_{a,\infty}$ and $\Gamma_b / \Gamma_{a,\infty}$. In (b), the reduced chemical potentials are defined as $\mu_a / k_{\rm B} T$ and $\mu_b / k_{\rm B} T$. The arrows indicate the approximate locations of the cpc (left arrow) and cmc (right arrow).}
\label{Fig:Figure_9}
\end{figure}

In the dilute regime, comparing the contaminated to the uncontaminated system, we observe a significant lowering of the surface tension due to the presence of the contamination. It was stressed by Rusanov \cite{Rusanov_book_1993} that this lowering implies that the contaminant must be more surface active than the original surfactant. This is demonstrated in the example shown in Figure \ref{Fig:adsorption} where we plot the adsorption of surfactant (C$_{12}$E$_8$) and contaminant as a function of surfactant concentration. The adsorption of contaminant is initially quite low but peaks at a distinct concentration due to its higher surface activity. In most experimental systems discussed here, it is also observed that the surface tension of the contaminated system slopes increasingly more downward (see, for example, the results for C$_{12}$E$_8$ in Figure \ref{Fig:C12E8_mixture}). This is an indication that the contaminant molecule takes up {\bf less surface} area than the surfactant molecule (for C$_{12}$E$_8$, less by a factor $\beta\!=$ 5). Any quantitative description for the full concentration regime {\em has} to take this factor into account.

%
%

At some distinct concentration below the regular cmc, the surface tension either levels off (see, for example, the results for LSA in Figure \ref{Fig:LSA_mixture_micel}) or increases (all other systems) due to the formation of mixed micelles (pre-micelles). We have named this concentration the {\it critical pre-micelle concentration} (cpc). Even though the formation of pre-micelles was suggested before, their precise composition and size remained elusive. Here, we assumed that the pre-micelles have the {\bf same size} as the regular micelles that are formed at the cmc, but with some of the regular surfactant molecules replaced by contaminant. We feel that this assumption is validated by the agreement with the experimental results.

At concentrations above the cpc, the adsorption of contaminant drops whereas the adsorption of surfactant increases as shown in Figure \ref{Fig:adsorption}. This is the mechanism that is usually alluded to to explain the observed minimum in the surface tension. It was, however, already remarked by Reichenberg \cite{Reichenberg_1947} (see also refs. \cite{Rusanov_book_1993, Franses_1977}) that the surface tension minimum is rather related to the {\it sign reversal} of the derivative of the chemical potential with respect to the concentration. In the example shown in Figure \ref{Fig:chemical_potential}, it is indeed demonstrated that the chemical potential of the contaminant slopes downward beyond the cpc indicating that the concentration of free contaminant peaks at the cpc before decreasing to its ultimate plateau value. 

At surfactant concentrations approaching the cmc, the composition of the pre-micelles gradually changes to contain less contaminant as shown, for instance, in Figure \ref{Fig:micellar_distribution}. Finally, when the concentration reaches the cmc, it signals the sudden and rapid formation of regular micelles (without contaminant) quickly outnumbering the number of mixed micelles. The surface tension almost attains its uncontaminated value with a small, yet distinct, difference remaining due to the continued presence of a small fraction of mixed micelles. 

With regard to the composition of the mixed micelles, it turns out to be necessary to allow only micelles containing a minimum number $m_{\rm min}$ of surfactant molecules. This means that, effectively, we have included a term in the micellar free energy that is infinite when $m_a \!<\! m_{\rm min}$. Ideally, one would like to include a term in the free energy that disfavours mixed micelles below a certain size in a more natural way. We have tested the inclusion of a surfactant-contaminant mixing entropy term and showed that it works well for C$_{12}$E$_8$ and C$_{10}$E$_8$, removing the necessity of a cut-off even leading to better agreement with experiment. However, including such a mixing entropy term does not work well for the experiments by McBain and Wood \cite{McBain_1940a} on LSA. Why this is the case is unknown and remains a point of concern. In fact, it seems difficult to improve on the agreement between theory without mixing entropy and experiment for LSA shown as the solid black line in Figure \ref{Fig:LSA_mixture_micel}.

In this article we have focused on the situation of a contaminated surfactant system. The contaminant (dodecanol) does not form micelles on its own and it is present in a tiny amount yet somehow capable of significantly lowering the surface tension. A minimum in the surface tension results and we have attempted to provide a quantitative description particularly of this effect. It then turns out to be necessary to determine the parameters of the contaminant -- $K_b$, $x_0^b$, and $\beta$ -- from a fit of the surface tension of the surfactant-contaminant {\bf mixture}. The parameter $K_b$ is therefore rather connected to the adsorption of the contaminant dodecanol on a surface covered with surfactant than on a surface of pure water. Furthermore, although we have seen that for C$_{12}$E$_8$ and SDS the value of $\beta$ determined in the fit ($\beta \!=$ 5 and $\beta \!=$ 3.5, respectively) are in close agreement with their values estimated independently from the value for $\Gamma_{b,\infty} \!=$ 11.1 $\times$ 10$^{-6}$ mol/m$^2$ of pure dodecanol on water \cite{Lin_1999} ($\beta \!=$ 5.02 and $\beta \!=$ 3.48, respectively), such an agreement is lacking for the C$_{10}$E$_8$ system. From a fit of the C$_{10}$E$_8$-dodecanol mixture one finds $\beta \!=$ 1.75 whereas on the basis of the adsorption measurements of the pure components on water, one would expect a value of $\beta \!\approx$ 5.4. This might be an indication that in this case attractive forces between C$_{10}$E$_8$ and dodecanol play a role on the surface. Still, the difference with C$_{12}$E$_8$ is striking and one would not expect this effect to be so dissimilar between similar surfactant molecules.

Despite the fact that we have focused on the situation of a contaminated surfactant system, the theoretical framework provided in this article can {\bf also} be applied to the mixture of two ordinary surfactants that are both capable of forming micelles. The fitting parameters of the second surfactant ($K_b$, $x_0^b$, and $\Gamma_{b,\infty}$) can then all be determined independently from the Gibbs isotherm of the system of the pure second surfactant. No additional fitting is then required to describe the surfactant-surfactant mixture. As an example of such an analysis, we apply our framework to the classical results by Clint \cite{Clint_1975} for two non-ionic surfactant mixtures in the Appendix. The agreement with the experiments is rather promising although it leaves room for a more thorough analysis.

To conclude, it is clear that further experimental and simulational testing of the simple model presented here is necessary. Furthermore, there seems to be room for improvement, especially regarding modeling of the free energy of the mixed micelles. The new expressions for the surface tension in Eqs.(\ref{eq:sigma_mixture}) and Eq.(\ref{eq:sigma_ions_mixture}) are then certainly of use to compare with experiment.

\renewcommand{\theequation}{A.\arabic{equation}}
\setcounter{equation}{0}

\appendix
\section*{Appendix A. Statistical Thermodynamic derivation of the Langmuir model for a surfactant mixture}

We discuss the Statistical Thermodynamic derivation of the Langmuir model for the adsorption of two surfactant types (species $a$ en $b$). Surfactant interactions on the surface are regarded as purely repulsive and taken into account by limiting the total amount of available surface positions to a maximum number $N_{\rm max}$. We shall further assume that one species (species $a$) takes up more of the available surface area than the other (species $b$) by a factor of $\beta$. Furthermore, when a surfactant molecule adsorbs at the surface from a (reference) bulk solution, a certain adsorption energy $\Delta E_{\rm s,i}$ is associated with it ($i \!=\! a, b$). The grand canonical partition function $\Xi$ can then be explicitly written down as:
\begin{equation}
\Xi = \sum\limits_{N_a=0}^{N_{\rm max}} \, \sum\limits_{N_b=0}^{\beta (N_{\rm max}-N_a)} \; W(N_a,N_b) \,
e^{-(\Delta E_{\rm s,a} + \mu^{\circ}_{\rm a} - \mu_a) N_a / k_{\rm B} T} \, e^{-(\Delta E_{\rm s,b} + \mu^{\circ}_{\rm b} - \mu_b) N_b / k_{\rm B} T} \,,
\end{equation}
where $W(N_a,N_b)$ is the number of ways $N_a$ molecules of type $a$ and $N_b$ smaller molecules of type $b$ can be distributed over $N_{\rm max}$ available positions. Each position can host 1 molecule of species $a$ and up to $\beta$ molecules of species $b$. 

The order of the summation is such that for a given number of large particles (species $a$) on the surface, all the remaining slots $\beta (N_{\rm max} - N_a)$ are available to the smaller species (species $b$). This would not be the case if the order of the summation is reversed. The precise distribution of smaller particles (and not just their number) would then be of influence to the total area available to the large particles. This means that the resulting expression for the grand canonical partition function is strictly valid only for $\beta \geqslant$ 1.

Carrying out the summation, we find that
\begin{eqnarray}
\Xi &=& \sum\limits_{N_a=0}^{N_{\rm max}} \; \binom{N_a}{N_{\rm max}} \, x_a^{N_a} \,
    \sum\limits_{N_b=0}^{\beta (N_{\rm max}-N_a)} \binom{N_b}{\beta (N_{\rm max}-N_a)} \, x_b^{N_b} \nonumber \\
&=& \sum\limits_{N_a=0}^{N_{\rm max}} \; \binom{N_a}{N_{\rm max}} \, x_a^{N_a} \, (1 + x_b)^{\beta (N_{\rm max}-N_a)} \nonumber \\
&=& (1 + x_b)^{\beta N_{\rm max}} \, (1 + \frac{x_a}{(1+x_b)^{\beta}})^{N_{\rm max}} = (x_a + (1 + x_b)^{\beta})^{N_{\rm max}} \,,
\end{eqnarray}
with $x_a$ and $x_b$ defined as
\begin{equation}
x_a \equiv \exp \, [ \, (\mu_a - \mu^{\circ}_{\rm a} - \Delta E_{\rm s,a}) / k_{\rm B} T \, ] \hspace*{10pt} {\rm and} \hspace*{10pt}
x_b \equiv \exp \, [ \, (\mu_b - \mu^{\circ}_{\rm b} - \Delta E_{\rm s,b}) / k_{\rm B} T \, ] \,.
\end{equation}
This gives for the surface grand free energy
\begin{equation}
\Omega = - k_{\rm B} T \, \ln(\Xi) = - N_{\rm max} \,\, k_{\rm B} T \, \ln(x_a + (1 + x_b)^{\beta}) \,.
\end{equation}
The average number of surfactants adsorbed at the surface are obtained from the grand free energy by differentiation with respect to the chemical potential:
\begin{equation}
\frac{N_a}{N_{\rm max}} = \frac{x_a}{x_a + (1 + x_b)^{\beta}} \hspace*{20pt} {\rm and} \hspace*{20pt} \frac{N_b}{\beta N_{\rm max}} = \frac{x_b \, (1 + x_b)^{\beta-1}}{x_a + (1 + x_b)^{\beta}} \,.
\end{equation}
In a more continuous form, in which the (discrete) number of surfactants adsorbed $N_a$ and $N_b$ are replaced by the surfactant adsorptions $\Gamma_a$ and $\Gamma_b$, these results lead to the following Langmuir isotherms \cite{EL_2018, EL_2024}
\begin{equation}
\frac{\Gamma_a}{\Gamma_{a,\infty}} = \frac{x_a}{x_a + (1 + x_b)^{\beta}} \hspace*{20pt} {\rm and} \hspace*{20pt} \frac{\Gamma_b}{\Gamma_{b,\infty}} = \frac{x_b \, (1 + x_b)^{\beta-1}}{x_a + (1 + x_b)^{\beta}} \,.
\end{equation}
where we have used that $\beta \!=\! \Gamma_{b,\infty} / \Gamma_{a,\infty}$.

The surface grand free energy is essentially the contribution to the total surface tension $\sigma$ due to the presence of surfactants. In a continuous form this gives:
\begin{equation}
\sigma = \sigma_0 - k_{\rm B} T \,\, \Gamma_{a,\infty} \, \ln(x_a + (1 + x_b)^{\beta}) \,.
\end{equation}
It can be verified that by setting either $x_a$ or $x_b$ to zero, one recovers the usual expression for a single type surfactant system.

It is important to stress that this expression is derived under the condition that $\beta \geqslant$ 1, i.e. species $a$ is taken to be the species that takes up more of the available surface area than species $b$. This is also reflected by the fact that the expression is not invariant under interchanging species $a \leftrightarrow b$.

\section*{Appendix B. Surfactant-surfactant mixture}

\begin{figure}[ht]
\centering
\subfloat[Figure 2 in Clint \cite{Clint_1975}]{\includegraphics[width=0.48\textwidth]{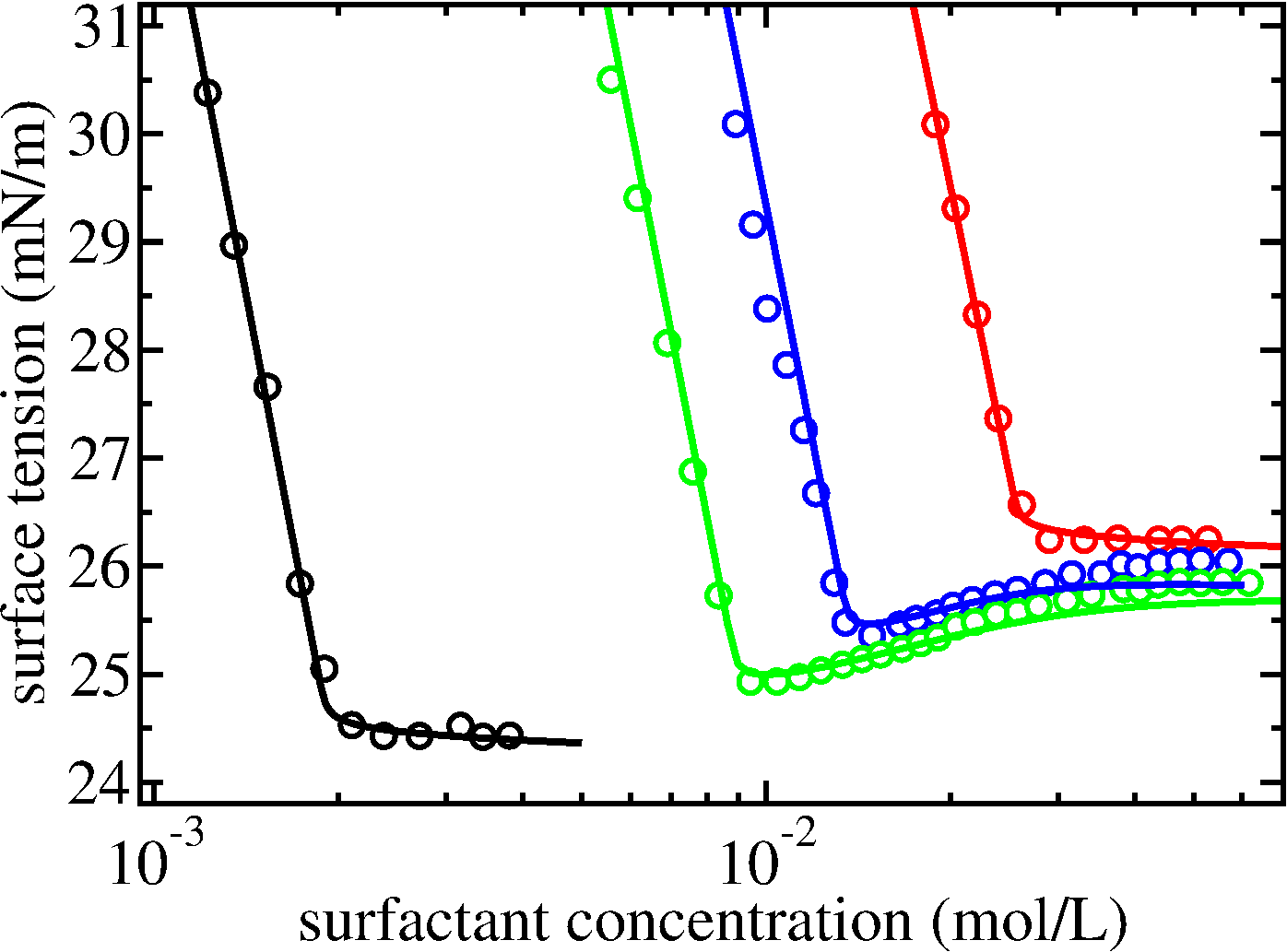}}
\hfill
\subfloat[Figure 6 in Clint \cite{Clint_1975}]{\includegraphics[width=0.48\textwidth]{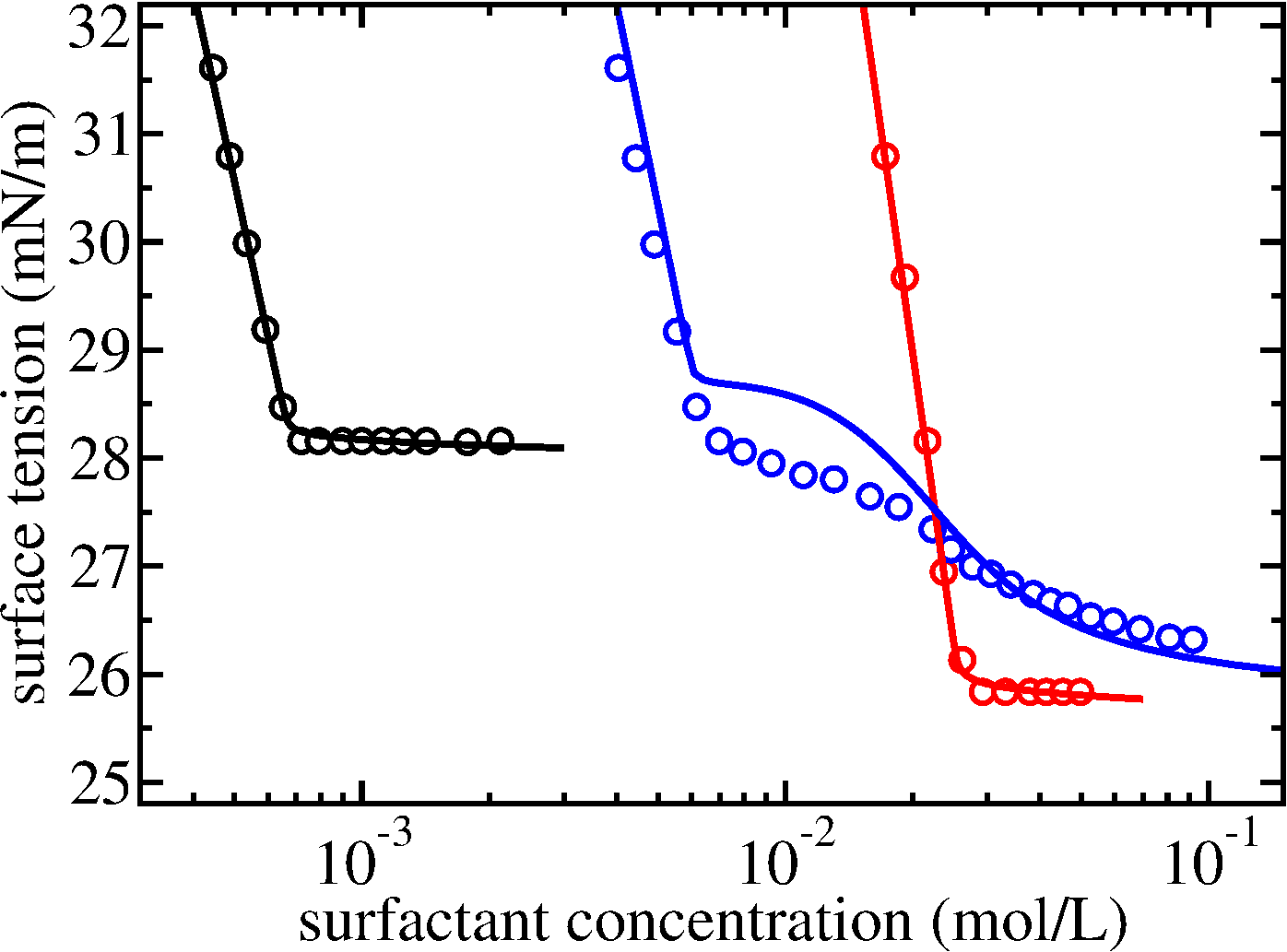}}
\caption{Surface tension as a function of surfactant concentration. In (a), symbols are the experimental results by Clint \cite{Clint_1975} for pure n-decyl methyl sulphoxide (black), pure n-octyl methyl sulphoxide (red) and for two mixtures with mole fraction of the n-decyl component $\alpha_m \!=$ 0.075 (blue) and $\alpha_m \!=$ 0.156 (green). In (b), symbols are the experimental results by Clint \cite{Clint_1975} for pure C$_{10}$E$_4$ (black), pure n-octyl methyl sulphoxide (red) and for a mixture with mole fraction of the C$_{10}$E$_4$ component $\alpha_m \!=$ 0.0892 (blue). The solid lines for the pure components (black and red curves) are determined from a fit with the fit parameters obtained listed in Table \ref{Table:Clint}. The solid lines for the mixtures (blue and green curves) are then determined without additional fitting. For the composition of the micelles we have assumed that $m_b = m - m_a$ and included mixing entropy. Furthermore, we have used that $\alpha = \alpha_m / (1 - \alpha_m)$.}
\label{Fig:Figure_10}
\end{figure}

The theoretical framework provided in this article can also be applied to the mixture of two ordinary surfactants that are both capable of forming micelles. The fitting parameters of the second surfactant ($K_b$, $x_0^b$, and $\Gamma_{b,\infty}$) can then all be determined independently from the Gibbs isotherm of the system of the pure second surfactant. No additional fitting is then required to describe the surfactant-surfactant mixture. As an example of such an analysis, we apply our framework to the classical results by Clint \cite{Clint_1975} for two non-ionic surfactant mixtures. Figure \ref{Fig:Figure_10} shows that agreement with the experiments is rather promising although it leaves room for a more thorough analysis.

%
%

%
%

\begin{table}[ht]
\centering
\begin{tabular}{l || c | c | c | c }
& $\Gamma_{\infty}$                               & $1/K$                                       & \hspace*{3pt} $m$ \hspace*{3pt} & $x_0$ \\
& \hspace*{1pt} 10$^{-6}$ mol/m$^2$ \hspace*{1pt} & \hspace*{1pt} 10$^{-6}$ mol/L \hspace*{1pt} &                                 & \hspace*{1pt} 10$^{-6}$ \hspace*{1pt} \\
\hline
\hline
n-decyl                         & 5.40  & 57.0 & 200 & 38.0  \\
n-octyl (Fig. 2 in Clint \cite{Clint_1975}) \hspace*{1pt} & 5.00  & 670  & 200 & 508   \\
\hline
C$_{10}$E$_4$                   & 3.20  & 2.70 & 200 & 13.35 \\
n-octyl (Fig. 6 in Clint \cite{Clint_1975})               & 5.00  & 640  & 200 & 502   \\
\end{tabular}
\caption{Values of the fit parameters used in Figure \ref{Fig:Figure_10} for the pure surfactants}
\label{Table:Clint}
\end{table}


\begin{thebibliography}{00}
\bibitem{reference_state}
It should be realized that $\Delta E_{\rm s}$ and $\Delta E_{\rm m}$ are defined as the energy gain for the surfactant molecule to become part of the surface or micelle, respectively, from a reference bulk solution. The absolute values calculated are therefore defined with respect to this reference state.
\end{thebibliography}
\end{document}